\documentclass[twocolumn,preprint]{aastex631}

\usepackage{amsmath}

\shorttitle{Extended stellar populations in UFDs}
\shortauthors{Tau et al.}

\begin{document}

\title{Extended stellar populations in Ultra-Faint Dwarf galaxies}

\correspondingauthor{Elisa A. Tau}
\email{elisa.tau@userena.cl}

\author[0000-0002-6925-221X]{Elisa A. Tau}
\affiliation{Departamento de Astronom{\'\i}a, Universidad de La Serena, Ra\'ul Bitr\'an 1305, La Serena, Chile}

\author[0000-0003-4341-6172]{A.~Katherina~Vivas}
\affiliation{Cerro Tololo Inter-American Observatory/NSF’s NOIRLab, Casilla 603, La Serena, Chile}

\author[0000-0002-9144-7726]{Clara E. Mart{\'\i}nez-V\'azquez}
\affiliation{Gemini Observatory/NSF’s NOIRLab, 670 N. A’ohoku Place, Hilo, HI 96720, USA}

\begin{abstract}

The possible existence of stellar halos in low-mass galaxies is being intensely discussed nowadays after some recent discoveries of stars located in the outskirts of dwarf galaxies of the Local Group. RR Lyrae stars can be used to identify the extent of these structures, taking advantage of the minimization of foreground contamination they provide. In this work we use RR Lyrae stars obtained from Gaia DR3, DES, ZTF, and Pan-STARRS1 to explore the outskirts of $45$ ultra-faint dwarf galaxies. We associate the stars with a host galaxy based on their angular separations, magnitudes and proper motions. We find a total of $120$ RR Lyrae stars that belong to $21$ different galaxies in our sample. We report seven new RR Lyrae stars in six ultra-faint dwarf galaxies (Hydrus I, Ursa Major I, Ursa Major II, Grus II, Eridanus II and Tucana II). We found a large number of new possible members in Bootes I and Bootes III as well, but some of them may actually belong to the nearby Sagittarius stream. Adding to our list of $120$ RR Lyrae stars the observations of other ultra-faint dwarf galaxies that were out of the reach of our search, we find that at least $10$ of these galaxies have RR Lyrae stars located at farther distances than $4$ times their respective half-light radius, which implies that at least $33\%$ of the 30 ultra-faint dwarfs with RR Lyrae star population have extended stellar populations.  

\end{abstract}

\keywords{galaxies: dwarf --- galaxies: halos --- galaxies: Local Group ---  galaxies: stellar content --- stars: variables: RR Lyrae stars}

\section{Introduction} \label{sec:intro}

Stellar halos of Milky Way-like galaxies have become objects of interest in the astronomical community and various studies regarding them have been carried out in these past years \cite[e.g.,][]{Zolotov2009, Cooper2010, Tissera2014, Peacock2015, Monachesi2016a, Monachesi2016b, Rey2022, Vera-Casanova2022}. These halos can be considered as a consequence of the hierarchical formation of structures of the $\Lambda$CDM model \citep{Searle1978, Frenk1988, Navarro1997} because stars can be found in the external regions of these massive galaxies principally due to the interactions between subhalos. On one hand, if the interacting subhalos are sufficiently massive in order to have a stellar component, they can deposit a significant fraction of this material in the outer regions of the host galaxies during the merging process \citep{Bullock2005, Monachesi2019}. On the other hand, during the interactions between galaxies, stars that were originally located in their inner regions can be heated up and end up reaching farther orbits \citep{Purcell2010, Tissera2013}.

However, when considering dwarf galaxies, the existence of stellar halos surrounding these low-mass structures is still a debatable subject. Until now the study of these extended stellar components in dwarf galaxies is limited on account of many reasons, the main one being that they would be extremely faint objects if they exist and hence they would be difficult to observe. Another difficulty relies on not having a well-modeled stellar mass-halo mass relation for these galaxies as we do in the cases of the Milky Way-like ones \citep{Behroozi2013, Moster2013}.

Nevertheless, recently there have been some studies that would favor the case in which these small galaxies also possess stellar halos. For example, \cite{Gilbert2022} study the dwarf galaxy M33 and conclude that it has a significant kinematically hot component that is likely the galaxy's stellar halo rather than a population from its thick disk. In addition, \cite{Qi2022} provide evidence for the presence of extended stellar halos or tidal debris in the outskirts of six classical dwarf spheroidals (dSphs), some of which also have RR Lyrae stars (RRL) outside of their King radii. More recently, \cite{Sestito2023a} found five stars with Gaia ERD3 making use of an algorithm presented in \cite{Jensen2023} and identified them as new members of Ursa Minor dSph, at elliptical distances ranging from $5.2$ to $11.7$ half-light radii ($R_h$). They also suggest the existence of a stellar halo around that galaxy, which would be more extended than previously thought. Furthermore, \cite{Sestito2023b} discovered two stars that belong to the Sculptor dSph which are located at $10\, R_h$  ($\sim 3$ kpc) from its center. Some studies focusing on the spatial distribution specifically of RRL in dSphs have also been undertaken. An example of this is the work of \cite{Stringer2021}, in which the authors found $51$ RRL uniformly distributed located beyond the tidal radius of Fornax. Another dSph found to have RRL located outside its tidal radius is Carina \citep{vivas13}, which suggests that the galaxy is undergoing a tidal disruption that led those stars to reach great distances. Additionally, \cite{Muraveva2020} studied $285$ RRL belonging to Draco with Gaia DR2 and found that those located in the southwestern region of this dSph seem to be closer to us, which could be indicating that Draco and the Milky Way are interacting.

Regarding ultra-faint dwarf galaxies (UFDs) specifically, there are also some recent studies that have searched for extended stellar populations. \citet{vivas20} identified extra-tidal RRL from Gaia DR2 in six UFDs (Bootes I, Bootes III, Sagittarius II, Reticulum III, Eridanus III, and Tucana III). Tucana II has been proven to have stars that reach up to $9\, R_h$ from its center, making it a very extended stellar system \citep{Chiti2021}. Hercules is also known for having stars located in its outer regions: \cite{garling18} identified $3$ RRL that lie outside its tidal radius, and more recently \cite{Longeard2023} reports $3$ new members of Hercules, one of them being at $9.5\, R_h$ from its center. Another example is Centaurus I, in which \cite{martinez21b} report a blue horizontal branch star at a distance $> 9 \, R_h$ and a RRL at $6 \, R_h$. Finally, in a recent work, \cite{Waller2023} report stars in the outskirts of Ursa Major I (one star at $3.7\, R_h$), Bootes I (one star at $4 R_h$) and Coma Berenices (two stars at $2.5\, R_h$). The authors suggest a tidal stripping scenario and the influence of supernovae feedback as the cause of these stars' locations in the three UFDs because they find that their chemical abundances are consistent with the central regions of their respective host galaxies. 

There have also been some recent studies from a theoretical point of view that analyze the possibility of the formation of stellar halos using numerical models. Two scenarios that have been proposed to affect the morphology of dwarf galaxies the most and that could lead to the formation of these stellar structures consist of mergers \citep{Deason2014} and tidal disruption \citep{Fattahi2018}. For example, \cite{Tarumi2021} analyze the possible formation of an UFD with an extended stellar distribution using cosmological hydrodynamic simulations and conclude that a major merger of two small galaxies can be responsible for this outcome. This mechanism is proposed as an explanation for the formation of the observed extended stellar halo of Tucana II. \cite{Deason2022} also use numerical models (N-body cosmological simulations) to argue in favor of the possible existence of stellar halos surrounding low-mass galaxies that could be formed due to this type of external process, and they analyze the properties of these extended structures differentiating between those stellar halos of satellite and those of field dwarf galaxies. In addition, \cite{Kado-Fong2022} use the FIRE-2 simulations to study the formation of extended and round stellar halos around dwarf galaxies and claim that these are built up by stars that formed in the galaxy (\textit{in-situ} stars) and then migrated to the external regions. Moreover, considering UFDs specifically, \cite{Ricotti2022} present an empirical model for the shape and mass of stellar halos of UFDs from cosmological N-body simulations. Their results were then compared to the observations of six isolated dwarf galaxies but they state that the systematic uncertainty of their inference on the star formation efficiency at the epoch of reionization is not negligible, which is mainly due to the lack of observations that extend to great distances from the center of these galaxies. Additionally, \cite{Goater2023} use the EDGE simulations to study the ellipticity and extended light surrounding tidally isolated UFDs and find that they exhibit a wide range of projected ellipticities, with many having anisotropic extended stellar halos that mimic tidal tails but are in fact a result of late-time accretion of lower mass stuctures. The authors also establish a connection between the final shape of the UFDs and their formation time.

Finding evidence of extended populations in UFDs is challenging due to the very low density of stars in these stellar systems, which is even more extreme in their external parts. Foreground stars dominate by large the field of view of the outskirts of the UFDs. Methods that have been used to observationally detect the stellar halos of UFDs described above include analyzing together several properties of the stars such as chemical abundances, kinematics, positions, proper motions and also their location in the color-magnitude diagram (CMD) from broad-band photometry. However, obtaining abundances and radial velocities for large number of stars in extended regions around UFDs is observationally expensive. 

In this work we follow \citet{vivas20} and use RRL as tracers of the stellar population of possible stellar halos in UFDs. The main advantage of using RRL is to avoid, or minimize, the foreground contamination. RRL are a type of variable star that are undergoing helium-burning phase in their core. They are old stars ($>10$ Gyr) and they trace a population that is ubiquitous in all UFD galaxies. Their pulsation periods are short (less than a day) and the amplitudes of their lightcurves is large \citep[$0.2-1.5$ mag in optical bands,][]{smith95}. RRL are considered standard candles \citep[e.g.,][]{Bono2003} due to the fact that they have a very well-defined relationship between their period of variability and their luminosity. This is a clear advantage when it comes to studying dwarf galaxies and its stellar halos: RRL can be associated with particular stellar systems based on their distance. Since the field population of Milky Way RRL decreases with Galactocentric radius, they become very rare at distances larger than $\sim 50$ kpc \citep[e.g.,][]{Vivas2006, zinn14, medina18, Stringer2021}. Most of the UFDs that are satellites of the Milky Way are located at larger distances and thus the possible contamination by halo field stars is usually negligible. Hence, the arduous task of excluding non-members of the UFDs that are part of the foreground population is considerably reduced.

Nowadays, thanks to new large-scale surveys that have been completed in the last few years, catalogs of RRL up to large distances in the Milky Way ($>80$ kpc) are readily available, including those coming from Gaia DR3 \citep{Clementini2022}, the Zwicky Transient Facility \cite[ZTF,][]{Huang2022}, the Dark Energy Survey \cite[DES,][]{Stringer2021}, and the Panoramic Survey Telescope and Rapid Response System \citep[Pan-STARRS,][]{sesar17}. The aim of this work is to search for RRL in these catalogs and find those associated to the UFD satellites of the Milky Way. A similar analysis was done by \cite{vivas20} based on Gaia DR2, but here we enhance it by {\sl (i)} using the updated catalogs mentioned above, which together cover uniformly the whole sky to distances beyond 100 kpc, {\sl (ii)} consistently searching the external regions of the UFDs in order to trace any possible extended population or halos around these galaxies, and {\sl (iii)} considering all UFDs in the Milky Way neighbourhood and not only those closer to 100 kpc. 

This paper is structured as follows: in \S~\ref{sec: data} we describe the catalogs that were used in this work. In \S~\ref{sec:methodology} we explain the methodology and the selection criteria to match RRL with UFDs. \S~\ref{sec:results} presents our results, and finally we discuss the conclusions in \S~\ref{sec:conclusions}.

\section{Data} \label{sec: data}

In this section we present the data used in this work. A brief description of the catalogs of RRL is given in \S~\ref{subsec:gaia} to \ref{subsec:PS1}, and in \S~\ref{subsec:UFDcatalog} we present the UFDs catalog that was used in this work.

\subsection{Gaia} \label{subsec:gaia}

Besides providing positional measurements for one billion stars in the Milky Way, the Gaia mission \citep{gaia16} collects multi-epoch photometry that allows the identification of variable stars such as RRL, down to a limiting magnitude of $G \approx 21$. \cite{vivas20} used the Gaia DR2 catalog of RRL stars \citep{clementini19} to search for stars belonging to UFDs. The DR2 catalog contained $140,784$ RRL stars covering unevenly the whole sky.  The mean completeness of that catalogue was $60\%$ but it had a strong dependence with position on the sky since not all the sky had been observed with the same frequency and light curves were too sparse to allow a reliable classification of variable stars in some parts of the sky. Consequently, it was not clear that the UFDs studied in \citet{vivas20} had indeed a complete census of RRL based in DR2.

With Gaia DR3 \citep{Gaia2022} the sky coverage became more uniform and the confirmation of RRL is more reliable. In addition, the errors of the proper motions provided by Gaia DR3 are smaller than those given by Gaia DR2, which makes it even more trustworthy. The DR3 catalog of RRL by \citet{Clementini2022} has $270,905$ stars, twice the number in DR2. By comparing with OGLE-IV \citep{soszynski19a, soszynski19b} in the Bulge and Magellanic Clouds regions, the completeness of the RRL catalog in DR3 is estimated to be between $82-94\%$, with contamination $<8\%$. The completeness at the faint end of Gaia DR3 is harder to asses since there are not many RRL stars known at such faint magnitudes, but by comparison with an all-sky compiled catalog (including stars in globular clusters and satellite galaxies), the completeness was found to be $\sim 64\%$ \citep{Clementini2022}.

The Gaia RRL DR3 catalog is the main source used in this work because it is the most complete catalog available in the literature, covering the whole sky, with complete light curves which allows a complete characterization of the RRL, and deeper than most of the other catalogs available. 

\subsection{ZTF} \label{subsec:ZTF}

ZTF is a time-domain survey being made with the Palomar $48$-inch Schmidt Telescope, and a $47$ sq deg field of view, $600$-megapixel camera \citep{Masci2019}. It scans the entire Northern sky ($\delta \geq -28\degr$) every two days. In this work we use a recent catalog by \cite{Huang2022} that has $71,755$ periodic variables from ZTF classified as RRL. The most distant RRL found in this catalog is located at $132$ kpc. Regarding its limiting magnitudes, the limits in the g, r and i bands are $20.8$ mag, $20.6$ mag and $19.9$ mag respectively. For heliocentric distances  $<80$ kpc, its completeness is $\gtrsim 80\%$. The depth of this survey is about the same as Gaia DR3.

\subsection{DES} \label{subsec:DES}

DES is an optical and near infrared survey \citep{flaugher15} using the Dark Energy Camera (DECam) at the 4m Blanco Telescope at Cerro Tololo Inter-American Observatory in Chile. The survey covers $5,000$ sq deg in the Southern Hemisphere with limiting magnitudes in the g, r, i, z and y bands of $24.3$, $24.1$, $23.5$, $22.9$, and $21.5$ mag, respectively. Although DES is not a variability survey, \cite{Stringer2021} used the few repeated observations of the same field over the 6-year survey to statistically identify RRL candidates.  \cite{Stringer2021} reports $6,971$ RRL candidates that are located at distances up to $\sim 335$ kpc and are all type RRab. The sample of sources provided by this catalog is estimated to be more than $70\%$ complete at distances of $\sim 150$ kpc. This survey is the deepest one of all of those used in this work. 

\subsection{Pan-STARRS1} \label{subsec:PS1}

The Panoramic Survey Telescope and Rapid Response System (Pan-STARRS) survey \citep{Chambers2016} is an optical and near infrared survey that covers all the sky north to a $- 30\degr$ declination. The first phase of the survey, called Pan-STARRS1 (PS1), was made with a 1.8m telescope at Haleakala Observatories near the summit of Haleakala in Hawaii. PS1 has only sparse multi-epoch data, with $\sim 12$ epochs in each of the five bands, taken over a 4.5-year period. \citet{sesar17} used template fitting and machine learning techniques to identify $\sim 45,000$ RRLs candidates in PS1 with $90\%$ purity and $80\%$ completeness down to $r_{P1} \sim 20$, or distance to the Sun of $\sim 80$ kpc. Although not as deep and complete as Gaia DR3, this catalogue may provide additional RRL for UFDs located north of $\delta=-30\degr$.

\bigskip
\bigskip

We bear in mind that PS1 and DES catalogs are based on few epochs, and some degree of contamination (low purity) is expected.  
Therefore, the stars selected from these catalogs will be considered as RRL candidates in this work.

Each of the catalogs used in this work was paired with Gaia DR3 to obtain proper motions and G magnitudes. The latter is important in order to have a common magnitude system for all stars. However, this could not be done for the DES catalog given that it contains a lot of stars that are fainter than those in Gaia DR3. In consequence, a transformation between magnitude systems was made. DES magnitudes of selected RRL (see Sec. \ref{subsec:angsep_select}) were transformed to G in a two-step procedure, first transforming to SDSS-based photometry using the transformation equations presented in DES DR2 \citep{Abbott2021}:
%Data Management website\footnote{\url{https://des.ncsa.illinois.edu/releases/dr2/dr2-docs/dr2-transformations}}:

\begin{equation}
g_{SDSS} = g_{DES} + 0.060 (g - i)_{DES} - 0.005
\label{eq:gDESStoSDSS}
\end{equation}

\begin{equation}
i_{SDSS} = i_{DES} + 0.167 (r - i)_{DES} - 0.027
\label{eq:iDESStoSDSS}
\end{equation}

Then, we obtained G mag for each RRL  using Eq.~\ref{eq:gSDSStoGaia} which was taken from \cite{Jordi2010}, with updated coefficients from \cite{Evans2018}:

%\begin{equation}
\begin{multline}
G = g_{SDSS} - 0.0940 - 0.5310(g - i)_{SDSS} \\
- 0.0974(g - i)^2_{SDSS} + 0.0052(g - i)^3_{SDSS}
\label{eq:gSDSStoGaia}
\end{multline}
%\end{equation}

\bigskip

\subsection{UFD catalog} \label{subsec:UFDcatalog}

Ultra-faint dwarf galaxies are the smallest structures known to be dark-matter-dominated up to now, along with being the oldest and most metal-poor systems \citep{simon19}. Over the past years, the discovery of local UFDs has been notoriously increasing due to new large-scale photometric surveys such as DES \citep{bechtol15,drlica15}. In this work we compiled a list of UFDs that are satellites of the Milky Way from Table 1 in \cite{vivas20} and Table A1 in \cite{martinez19}. In addition, we considered recent UFDs discovery reports, including Centaurus I \citep{Mau2020}, Eridanus IV \citep{Cerny2021b}, and Pegasus IV \citep{Cerny2022}. We also included DELVE 2 \citep{Cerny2021a} although it has not yet been confirmed as a UFD. The final sample of UFDs used in this work is listed in Table \ref{table:dataUFDs} and consists of $45$ galaxies. The properties of each galaxy are presented as follows: galaxy name, RA (J2000), declination (J2000); proper motion and proper motion errors, half-light radius in arcminutes, half-light radius in kiloparsecs, absolute magnitude in V ($M_v$), ellipticity ($\epsilon$), heliocentric distance and position angle.

The values presented in Table \ref{table:dataUFDs} were taken from \cite{Belokurov2007, Irwin2007, moretti09, Sand2009, mcconnachie12, torrealba16b, Medina2017, Martin2016, carlin18, li18, Homma2018, martinez19, vivas20, Mau2020, simon20, garofalo21, MartinezVazquez2021a, Cerny2021a, Cerny2021b, pace22} and \cite{Cerny2022}. However, the proper motion values were updated with those presented in \cite{mcconnachie2020} and \cite{Li2021}.

The order of magnitude of the stellar mass of the UFDs analyzed in this work ranges from $\sim 10^3 \, M_\odot$ to $10^6 \, M_\odot$ (e.g., M$^*_{\rm DELVE2} = 8.8 \times 10^2 \, M_\odot$ \citep{Cerny2021a}, M$^*_{\rm UMaI} = 1.4 \times 10^4 \, M_\odot$ \citep{mcconnachie12}, M$^*_{\rm TucIV} = 2.2 \times 10^6 \, M_\odot$ \cite[][]{simon20}).

\begin{deluxetable*}{lrrrrrrrrrrrl}
\rotate
\tabletypesize{\scriptsize}
\tablecolumns{13}
\tablewidth{0pc}
\tablecaption{Parameters of the $45$ UFDs studied in this work. \label{table:dataUFDs}}
\tablehead{
Galaxy &  RA &  DEC & $\mu_{\alpha}  \cos(\delta)$ &  $\sigma \mu_{\alpha} \cos(\delta)$ &  $\mu_\delta$ &  $\sigma \mu_\delta$ & $R_h$ & $R_h$ & $M_V$ &  $\epsilon$ &  Distance &  PA \\
& [deg] & [deg] & [mas/yr] & [mas/yr] & [mas/yr] & [mas/yr] & [$\arcmin$] & [kpc] & [mag] & & [kpc] & ($\degr$)  \\
        }
\startdata
Draco II &  238.1740 &    64.5790 & 1.011 &  0.115 & 0.956 &  0.126 & 3.00 & 19.00 &  -0.80 &  0.23 &  21.50 &   76.00  \\
       Tucana III &  359.1075 &   -59.5831 &                                 -0.111 &                                              0.023 &                 -1.629 &                                  0.023 &       5.10 &       33.97 &        -1.30 &        0.20 &           22.90 &   25.00 \\
          Segue I &  151.7504 &    16.0756 &                                 -2.074 &                                              0.052 &                 -3.411 &                                  0.043 &       3.93 &       29.00 &        -1.30 &        0.32 &           23.00 &   75.00                                             \\
         Hydrus I &   37.3890 &   -79.3089 &                                  3.775 &                                              0.027 &                 -1.513 &                                  0.027 &       6.60 &       53.00 &        -4.71 &        0.20 &           27.60 &   97.00                                                 \\
       Carina III &  114.6298 &   -57.8997 &                                  3.082 &                                              0.068 &                  1.394 &                                  0.072 &       3.75 &       30.00 &        -2.40 &        0.55 &           27.80 &  150.00                                             \\
    Triangulum II &   33.3225 &    36.1719 &                                  0.602 &                                              0.081 &                  0.085 &                                  0.093 &       2.50 &       32.00 &        -4.20 &        0.30 &           28.40 &   73.00                                          \\
         Cetus II &   19.4700 &   -17.4200 &                                  2.840 &                                              0.060 &                  0.460 &                                  0.065 &       1.90 &       16.58 &         0.00 &        0.00$^*$ &           30.00 &    0.00$^*$ \\
     Reticulum II &   53.9203 &   -54.0513 &                                  2.391 &                                              0.029 &                 -1.379 &                                  0.032 &       5.41 &       32.00 &        -3.88 &        0.56 &           30.00 &   69.00                                          \\
    Ursa Major II &  132.8726 &    63.1335 &                                  1.701 &                                              0.120 &                 -1.845 &                                  0.128 &      13.90 &      149.00 &        -4.25 &        0.55 &           32.00 &  -76.00                \\
         Segue II &   34.8226 &    20.1624 &                                  1.425 &                                              0.061 &                 -0.313 &                                  0.052 &       3.64 &       35.00 &        -1.86 &        0.21 &           35.00 &  166.00                                         \\
        Carina II &  114.1066 &   -57.9991 &                                  1.887 &                                              0.043 &                  0.164 &                                  0.043 &       8.69 &       91.00 &        -4.50 &        0.34 &           37.40 &  170.00                               \\
        Willman I &  162.3436 &    51.0501 &                                  0.295 &                                              0.086 &                 -1.074 &                                  0.131 &       2.52 &       25.00 &        -2.53 &        0.47 &           38.00 &   74.00                                  \\
        Bootes II &  209.5141 &    12.8553 &                                 -2.273 &                                              0.151 &                 -0.361 &                                  0.115 &       3.07 &       51.00 &        -2.94 &        0.23 &           42.00 &  -71.00                                 \\
   Coma Berenices &  186.7454 &    23.9069 &                                  0.374 &                                              0.059 &                 -1.699 &                                  0.056 &       5.67 &       77.00 &        -4.38 &        0.37 &           44.00 &  -58.00                                            \\
       Bootes III &  209.3000 &    26.8000 &                                 -1.176 &                                              0.019 &                 -0.890 &                                  0.015 &      33.03 &      446.79 &        -5.75 &        0.33 &           46.50 &  -81.00                                  \\
        Tucana IV &    0.7170 &   -60.8300 &                                  0.618 &                                              0.036 &                 -1.696 &                                  0.038 &       9.30 &      127.00 &        -3.00 &        0.39 &           47.00 &   27.00                               \\
         Tucana V &  354.3500 &   -63.2700 &                                 -0.270 &                                              0.083 &                 -1.253 &                                  0.110 &       1.00 &       34.00 &        -1.60 &        0.70 &           55.00 &   30.00                                       \\
          Grus II &  331.0200 &   -46.4400 &                                  0.389 &                                              0.034 &                 -1.526 &                                  0.036 &       6.00 &       94.00 &        -3.90 &        0.20$^\dagger$ &           55.00 &    0.00$^*$ \\
        Tucana II &  343.0600 &   -58.5700 &                                  0.935 &                                              0.031 &                 -1.243 &                                  0.036 &       7.20 &      165.00 &        -3.90 &        0.00$^*$ &           58.00 &    0.00$^*$ \\
         Bootes I &  210.0200 &    14.5135 &                                 -0.307 &                                              0.052 &                 -1.157 &                                  0.043 &      10.50 &      242.00 &        -6.02 &        0.26 &           66.00 &    6.00                              \\
         Delve II &   28.7720 &   -68.2530 &                                  1.020 &                                              0.240 &                 -0.850 &                                  0.180 &       1.02 &       21.00 &        -2.10 &        0.03 &           71.00 &   74.00                               \\
   Sagittarius II &  298.1663 &   -22.0650 &                                 -0.710 &                                              0.077 &                 -0.905 &                                  0.051 &       1.70 &       35.00 &        -5.70 &        0.00 &           73.10 &  103.00                                       \\
      Eridanus IV &   76.4380 &    -9.5150 &                                  0.250 &                                              0.060 &                 -0.100 &                                  0.050 &       3.30 &       75.00 &        -4.70 &        0.54 &           76.70 &   65.00                                      \\
    Horologium II &   49.1077 &   -50.0486 &                                  0.915 &                                              0.330 &                 -0.913 &                                  0.439 &       2.17 &       47.00 &        -1.56 &        0.71 &           78.00 &  137.00                                 \\
     Horologium I &   43.8813 &   -54.1160 &                                  0.865 &                                              0.047 &                 -0.601 &                                  0.048 &       1.17 &       30.00 &        -3.55 &        0.32 &           79.00 &   53.00                                      \\
          Virgo I &  180.0400 &    -0.6800 &                                 -0.320 &                                              0.140 &                 -0.620 &                                  0.090 &       1.50 &       38.00 &        -0.80 &        0.44 &           87.00 &   51.00                                \\
     Eridanus III &   35.6952 &   -52.2838 &                                  1.080 &                                              0.295 &                 -0.490 &                                  0.150 &       0.34 &        8.60 &        -2.37 &        0.57 &           87.00 &   73.00                             \\
       Pegasus IV &  328.5390 &    26.6200 &                                  0.330 &                                              0.070 &                 -0.210 &                                  0.080 &       1.55 &       41.00 &        -4.25 &        0.41$^\dagger$ &           90.00 &  115.00 \\
    Reticulum III &   56.3600 &   -60.4500 &                                  0.519 &                                              0.222 &                 -0.173 &                                  0.247 &       2.40 &       64.23 &        -3.30 &        0.00$^*$ &           92.00 &    0.00$^*$ \\
     Ursa Major I &  158.7706 &    51.9479 &                                 -0.387 &                                              0.058 &                 -0.641 &                                  0.068 &       8.13 &      319.00 &        -5.12 &        0.59 &           97.00 &   67.00                     \\
       Phoenix II &  354.9928 &   -54.4050 &                                  0.501 &                                              0.062 &                 -1.199 &                                  0.076 &       1.50 &       43.63 &        -2.70 &        0.40 &          100.00 &  156.00                             \\
      Aquarius II &  338.4813 &    -9.3274 &                                  0.647 &                                              0.588 &                 -0.298 &                                  0.548 &       5.10 &      159.00 &        -4.30 &        0.40 &          107.87 &  121.00                                  \\
      Centaurus I &  189.5850 &   -40.9020 &                                 -0.140 &                                              0.050 &                 -0.200 &                                  0.040 &       2.30 &       79.00 &        -5.55 &        0.40 &          116.30 &   20.00                              \\
           Grus I &  344.1760 &   -50.1630 &                                  0.071 &                                              0.056 &                 -0.270 &                                  0.078 &       1.77 &       62.00 &        -3.40 &        0.41 &          127.00 &    4.00                                   \\
         Hercules &  247.7583 &    12.7917 &                                 -0.030 &                                              0.040 &                 -0.360 &                                  0.030 &       6.27 &      330.00 &        -5.80 &        0.67 &          133.00 &  -72.59                          \\
         Hydra II &  185.4254 &   -31.9853 &                                 -0.576 &                                              0.280 &                 -0.101 &                                  0.212 &       1.70 &       66.00 &        -4.60 &        0.00$^*$ &          151.00 &    0.00$^*$  \\
           Leo IV &  173.2375 &    -0.5333 &                                  0.007 &                                              0.174 &                 -0.261 &                                  0.134 &       3.30 &      206.00 &        -5.00 &        0.25 &          154.00 &  355.00                                   \\
Canes Venatici II &  194.2917 &    34.3208 &                                 -0.138 &                                              0.111 &                 -0.320 &                                  0.082 &       1.85 &       74.00 &        -5.20 &        0.23 &          159.00 &    9.45                        \\
            Leo V &  172.7900 &     2.2200 &                                  0.118 &                                              0.212 &                 -0.387 &                                  0.151 &       1.00 &      135.00 &        -4.40 &        0.43 &          173.00 &  -71.00                                          \\
      Pegasus III &  336.1000 &     5.4050 &                                  0.060 &                                              0.100 &                 -0.200 &                                  0.100 &       0.85 &       53.00 &        -3.40 &        0.38 &          174.00 &    0.00$^*$ \\
        Pisces II &  344.6292 &     5.9525 &                                  0.675 &                                              0.299 &                 -0.631 &                                  0.212 &       1.12 &       58.00 &        -4.10 &        0.31$^\dagger$ &          175.00 &    0.00$^*$ \\
        Columba I &   82.8542 &   -28.0425 &                                  0.189 &                                              0.119 &                 -0.556 &                                  0.134 &       2.20 &      117.11 &        -4.20 &        0.30 &          183.00 &   24.00                               \\
        Cetus III &   31.3310 &    -4.2700 &                                  0.140 &                                              0.080 &                 -0.160 &                                  0.080 &       1.23 &       89.81 &        -2.45 &        0.76 &          251.00 &  101.00                                          \\
      Eridanus II &   56.0880 &   -43.5334 &                                  0.210 &                                              0.090 &                 -0.050 &                                  0.110 &       2.30 &      277.00 &        -7.10 &        0.40 &          370.00 &   22.84                                 \\
            Leo T &  143.7225 &    17.0514 &                                 -0.010 &                                              0.050 &                 -0.110 &                          0.050 &       1.40 &      120.00 &        -7.60 &        $\sim$0.00 &         409.00 &    $\sim$0.00 \\
\enddata
\tablecomments{$^*$: values are not measured; $^\dagger$: values are upper limits.}
\end{deluxetable*} 

\section{Methodology} \label{sec:methodology}

We associate RRL with UFD galaxies in three steps: {\sl (i)} angular separation, {\sl (ii)} heliocentric distance, and {\sl (iii)} proper motions. The selection constraints are explained in Secs. \ref{subsec:angsep_select}, \ref{subsec:mag_select} and \ref{subsec:pm_select} and were applied separately to each catalog used in this work.

\subsection{Angular separation selection} \label{subsec:angsep_select}

The first condition we took into account was that the stars should be located no farther than a certain radius from the center of the galaxy to be considered as a member candidate. To consistently search in the external parts of UFD galaxies, we set up the maximum search radius as a function of the $R_h$ of each galaxy. The maximum separation considered between the stars and the center of the galaxy was $15 R_h$, except for very small galaxies (i.e., if $15 R_h<1\degr$) where we allowed a search radius of $1\degr$. This exception was prompted by previous results by \citet{vivas20} where the authors found possible distant members in very small galaxies like Sagittarius II or Eridanus III (which have $R_h$ of $1\farcm 7$ and $0\farcm 34$, respectively). 

\subsection{Magnitude selection} \label{subsec:mag_select}

After having made a first cut out considering the separation between the stars and the center of the UFD, we proceeded to enhance our selection by selecting only RRL that are at the same distance as the UFD. To do so, we considered the UFD's distances (column $11$ in Table \ref{table:dataUFDs}) to estimate the magnitude $m_G$ that the RRL would have in Gaia's G-band if they belong to the galaxy using the distance modulus equation: 

\begin{equation}
m_G - M_G = 5 \cdot \log(d) - 5
\label{eq:distancemodulus}
\end{equation}

\noindent
where the distance d is the distance to the UFD in pc. The absolute magnitude of an RRL, $M_G$, was assumed to be 0.46, a value obtained using the \citet{muraveva18}'s $M_G$ - [Fe/H] relationship ($M_G = 0.32 {\rm [Fe/H]} + 1.1$), and assuming a mean metallicity of [Fe/H]$=-2$ dex, which is typical for UFDs \citep{simon19}. Since not all UFDs have measured metallicities, we decided to use [Fe/H]=-2.0 as a reference value. However, the range of metallicities in UFDs is not large, and translates in maximum differences in $M_G$ of only 0.25 mag.

The expected magnitude of RRL found this way are de-reddened. Thus, we proceed to correct for extinction the $G$ magnitudes of the RRL selected in the previous step. To do that, we computed the color excess E(B-V) \citep{schlegel98} using the \textsc{Python} package \textit{dustmaps} \citep{green18}. We obtained the reddening correction ($A_G$) using the following relation: $A_G = 2.740 \cdot E(B-V)$ \citep{casgrande18}.  

\begin{figure*}[!ht]
    \begin{centering}
    \includegraphics[width=\columnwidth]{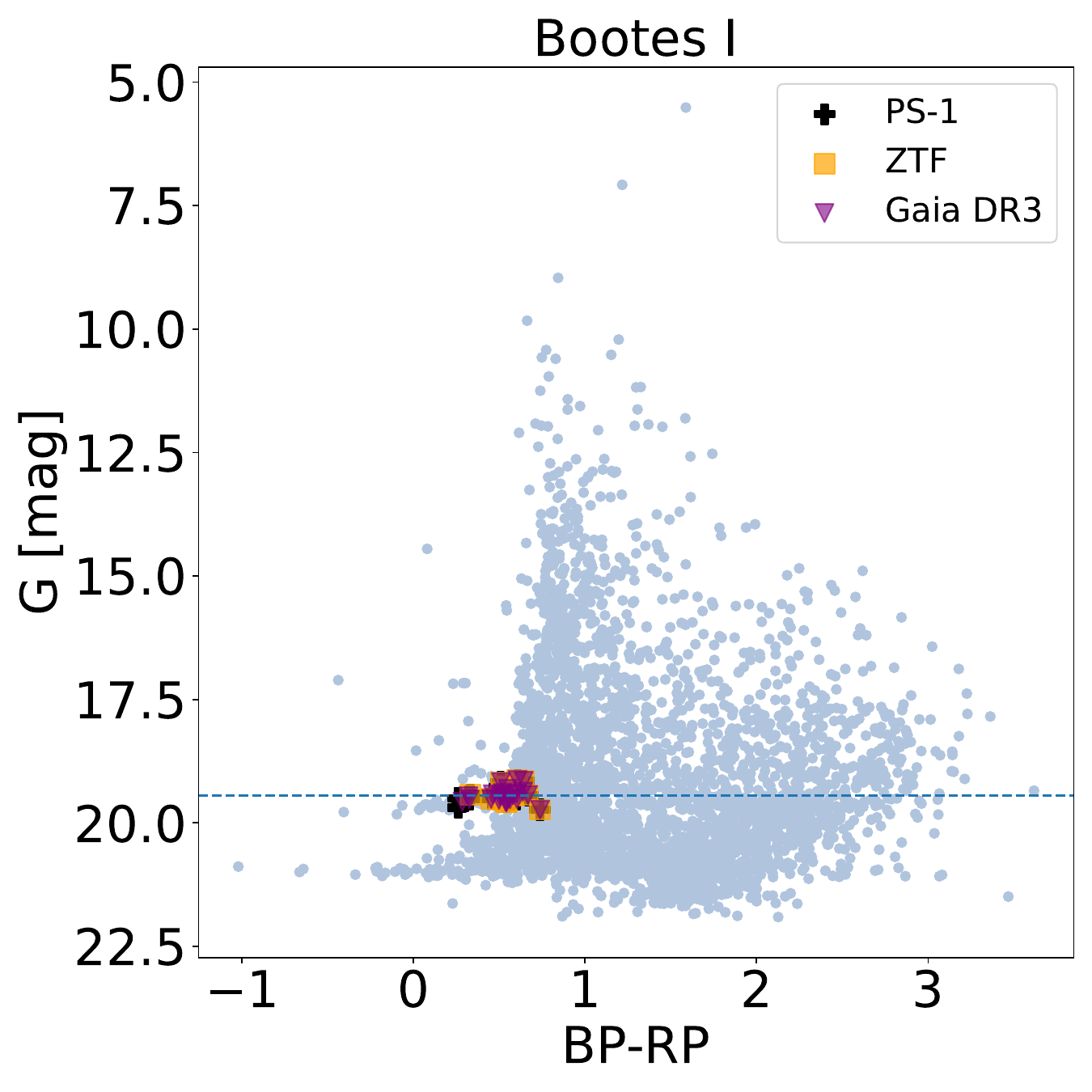}
    \includegraphics[width=\columnwidth]{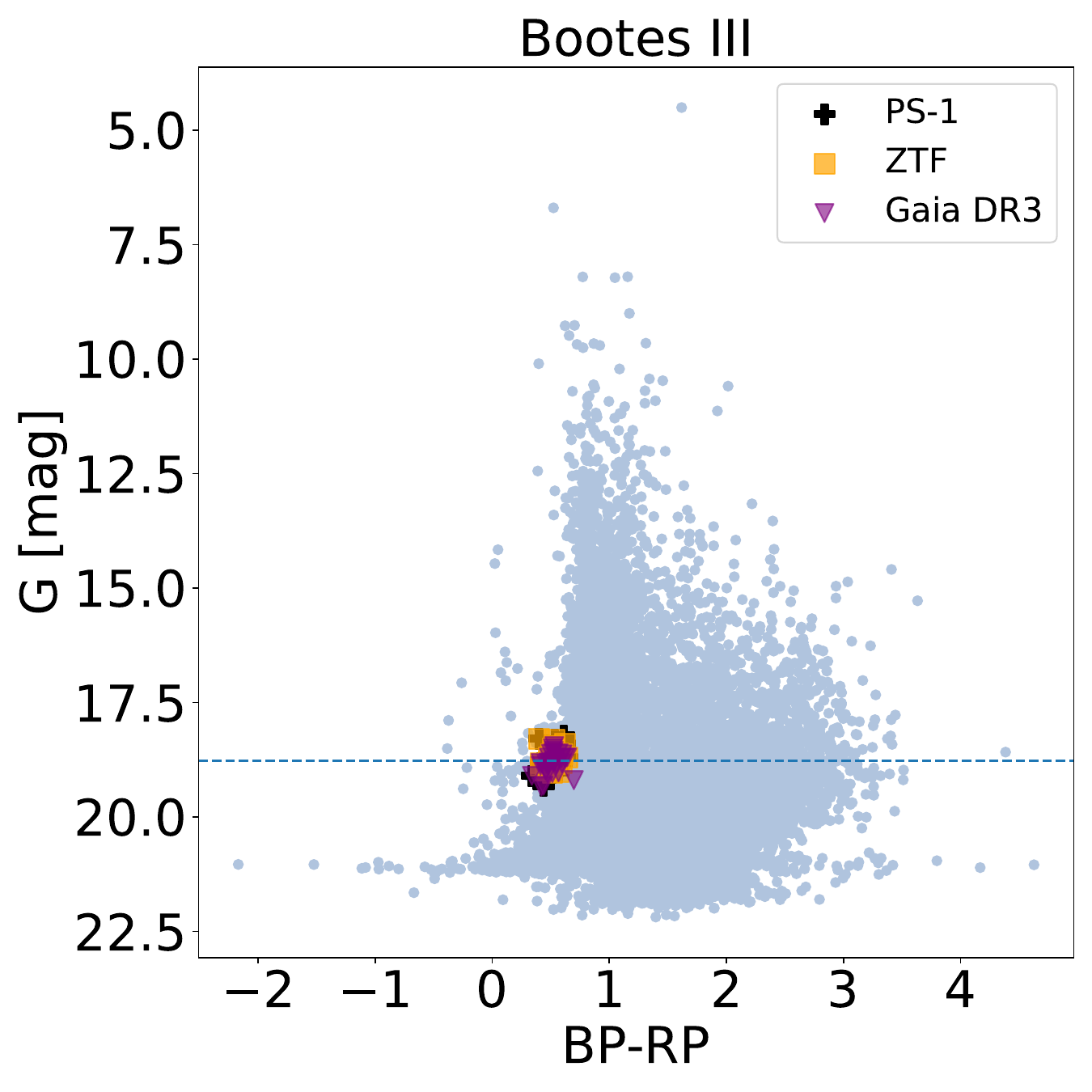}
    \includegraphics[width=\columnwidth]{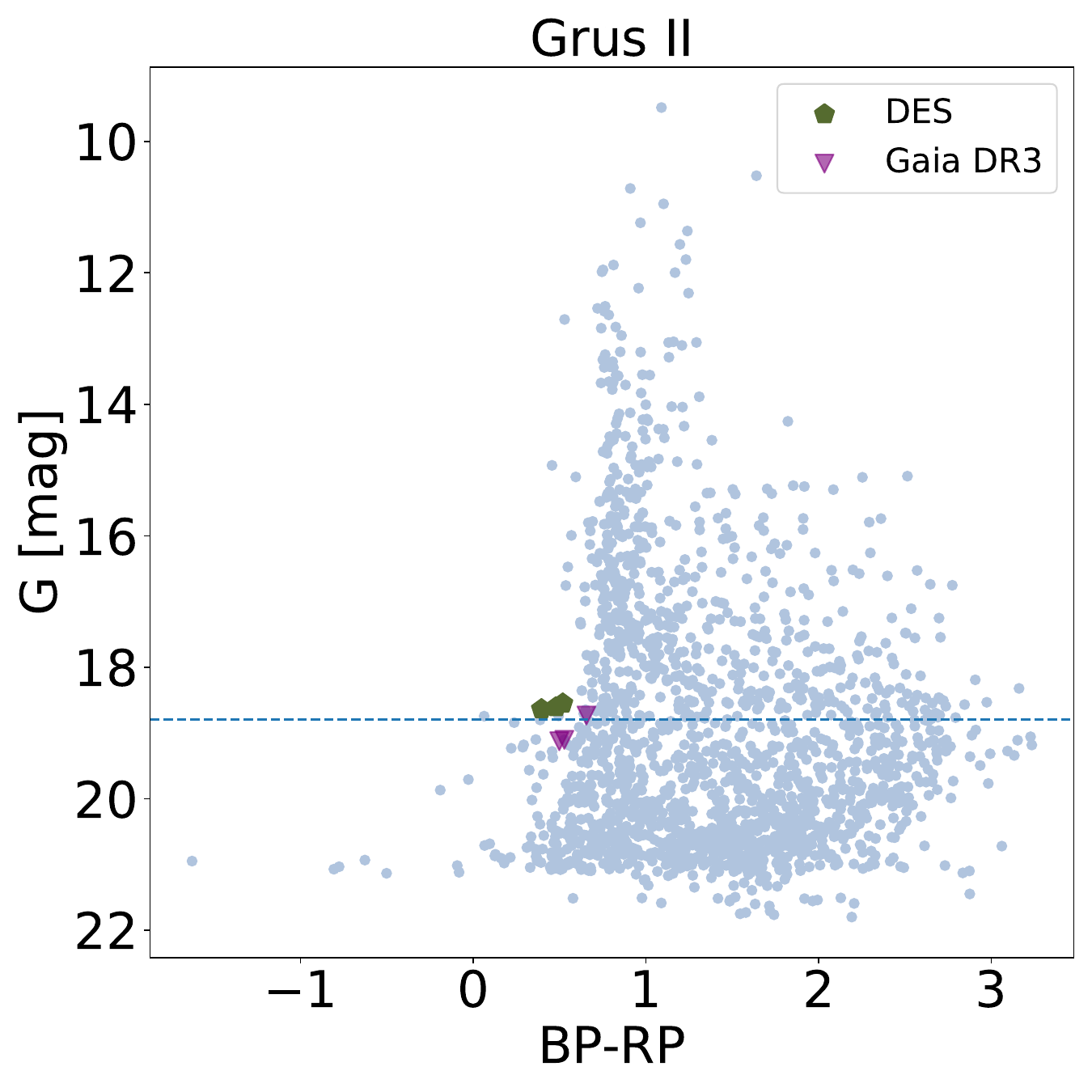}
    \includegraphics[width=\columnwidth]{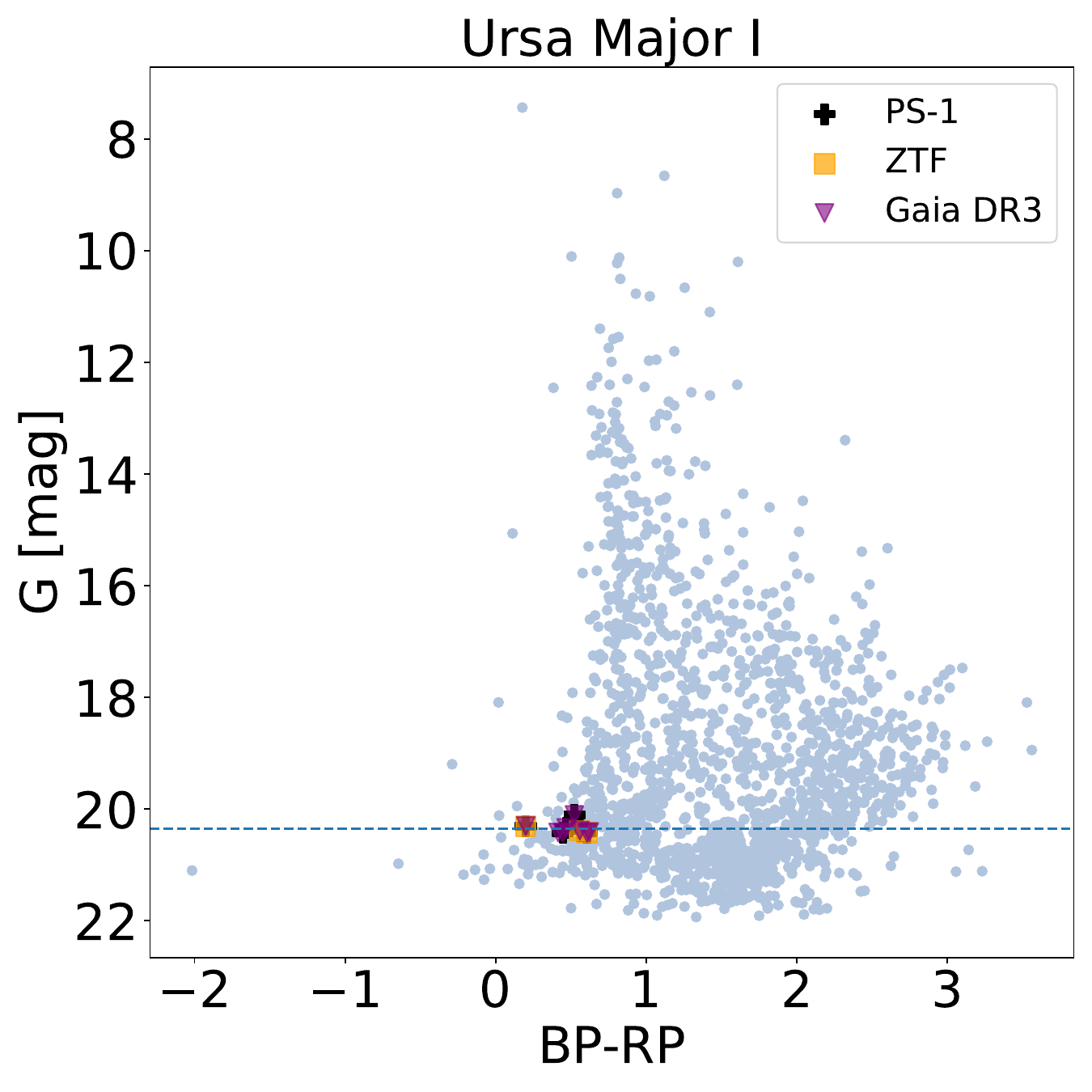}
    \caption{Gaia CMDs showing four galaxies of our sample. All stars enclosed within $3 \, R_h$ are plotted in the background, except for Bootes III for which only those stars enclosed within $1\deg$ are shown. The RRL selected for each galaxy are located in the horizontal branch as expected. The horizontal dashed line indicates the G mean value of the selected RRL. In the case of Grus II, a double horizontal branch is seen (see discussion in the text). The different symbols represent the catalog where the RRL were selected from. \label{fig:GvsBPRP}}
    \end{centering}
\end{figure*}

Finally, we selected RRL having extinction-corrected magnitudes within $\pm 0.5$ mag of the expected magnitude for each UFD. The range allows for differences arising from different metallicities, photometric errors, and/or small distance gradients within the searched region. An example of the selection after this step can be seen in Fig. \ref{fig:GvsBPRP} for four of the galaxies in our sample.

This cut in magnitude eliminates most foreground/background RRL which will be brighter/fainter than the RRL in the UFD. The cut also eliminates any possible contamination by Anomalous Cepheids, whose periods overlap with RRL and have similar lightcurve shapes. Anomalous Cepheids are however $\sim 0.5–2.5$ mag brighter than RRL \citep{catelan15} and hence will be naturally excluded with the magnitude cut we applied.

\subsection{Proper motion selection} \label{subsec:pm_select}

Finally, we considered the proper motion of the previously selected RRL candidates and the UFDs to dismiss any possible non-member stars that had passed the other two selection criteria mentioned in Secs. \ref{subsec:angsep_select} and \ref{subsec:mag_select}. 

\begin{figure*}[!ht]
    \begin{centering}
    \includegraphics[width=\columnwidth]{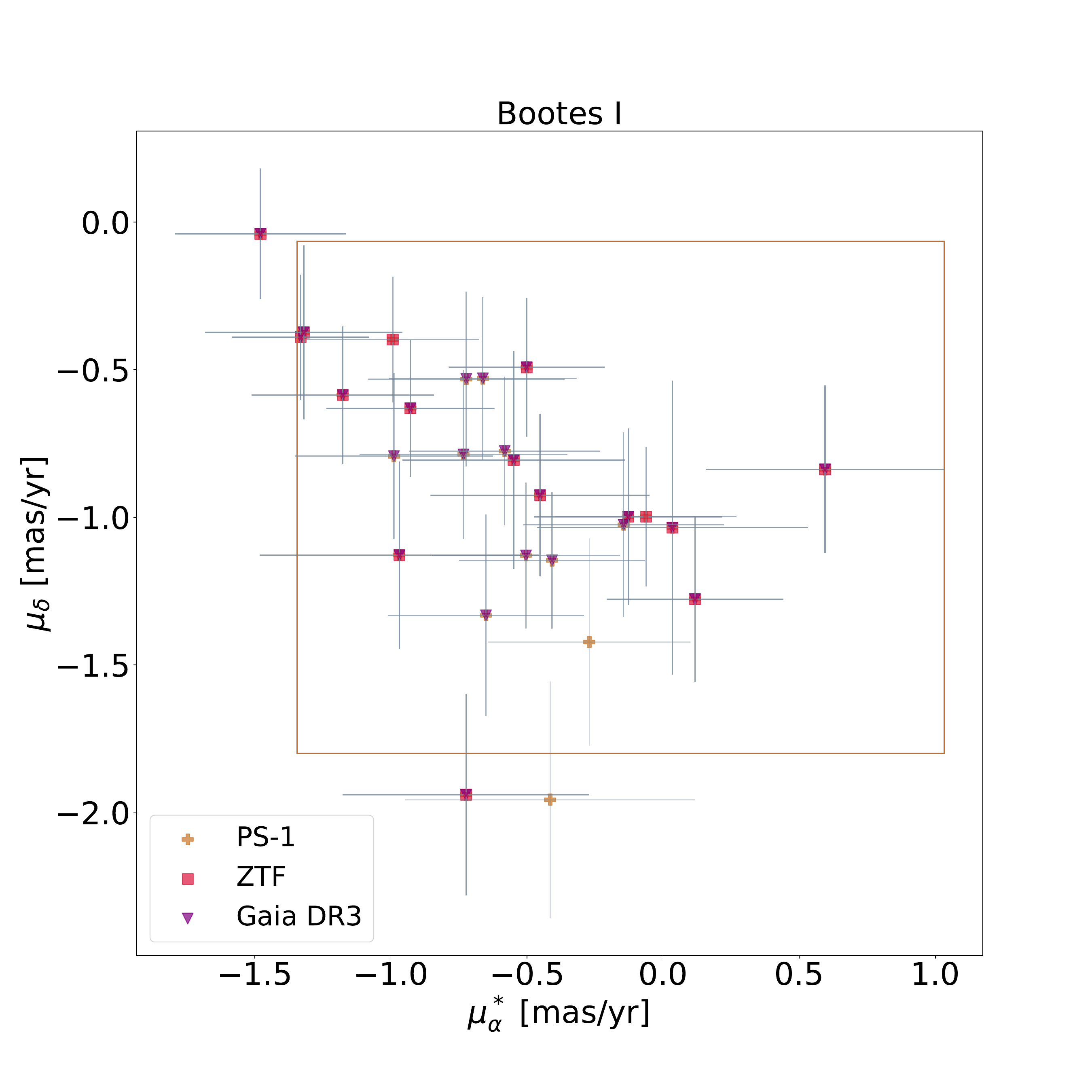}
    \includegraphics[width=\columnwidth]{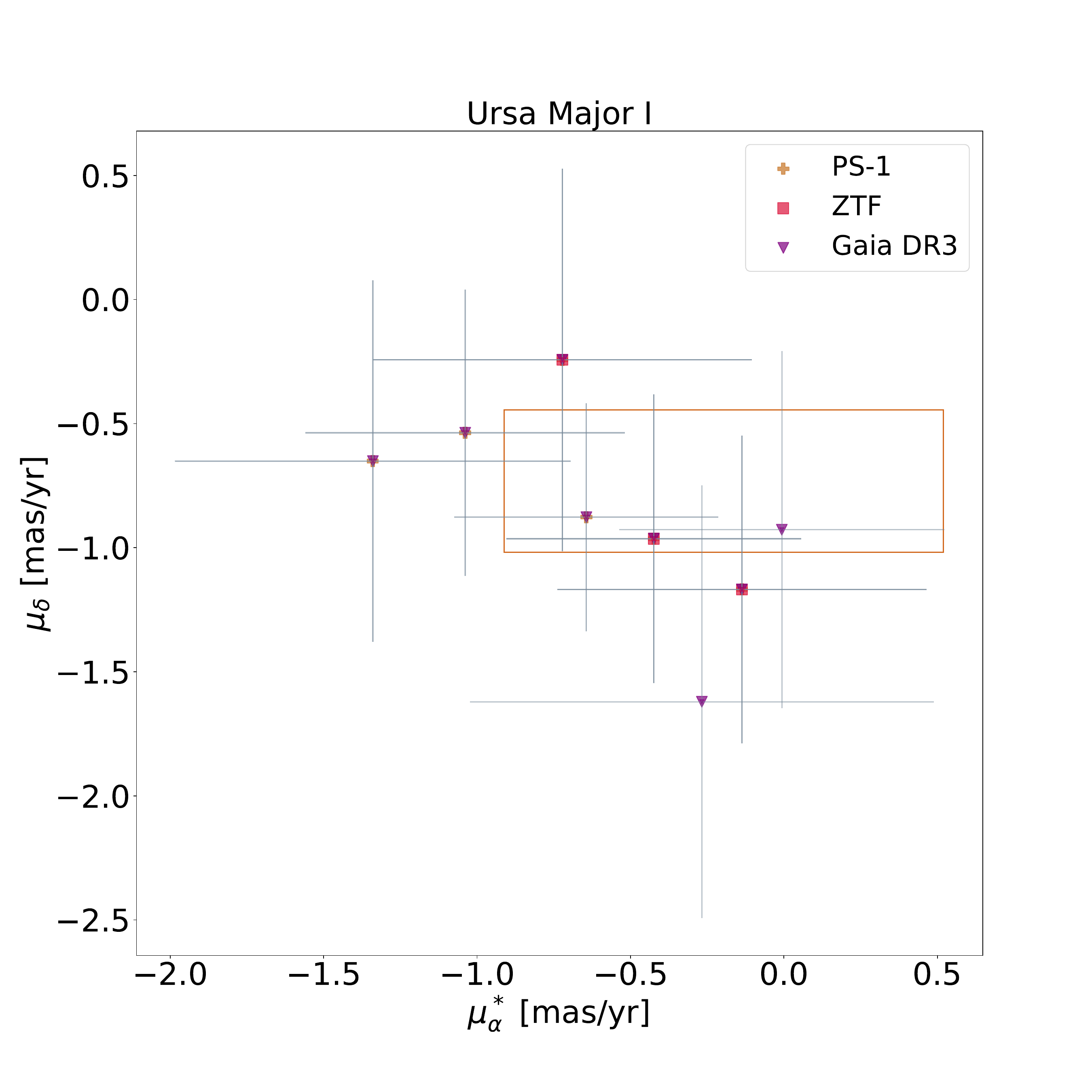}
    \includegraphics[width=\columnwidth]{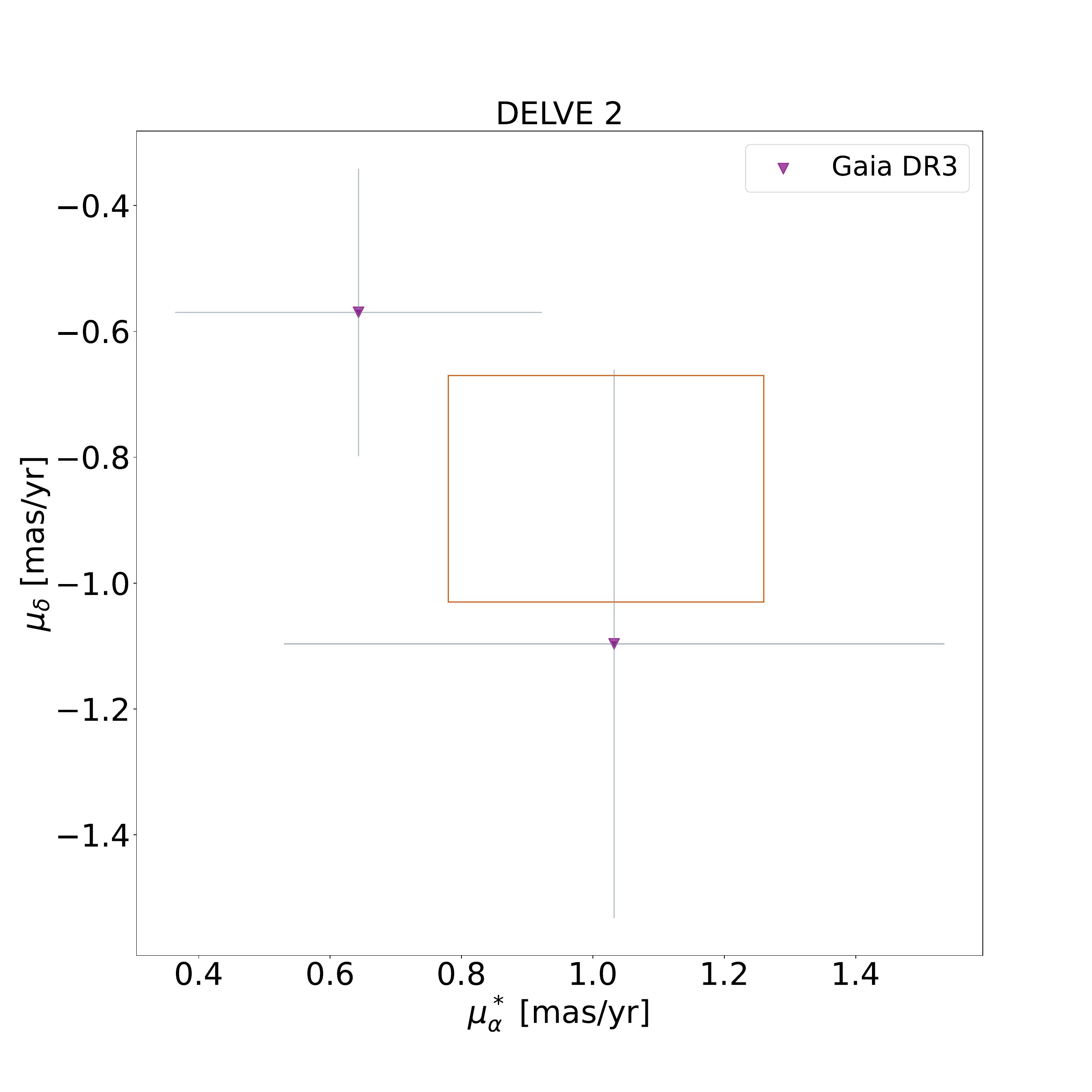}
    \includegraphics[width=\columnwidth]{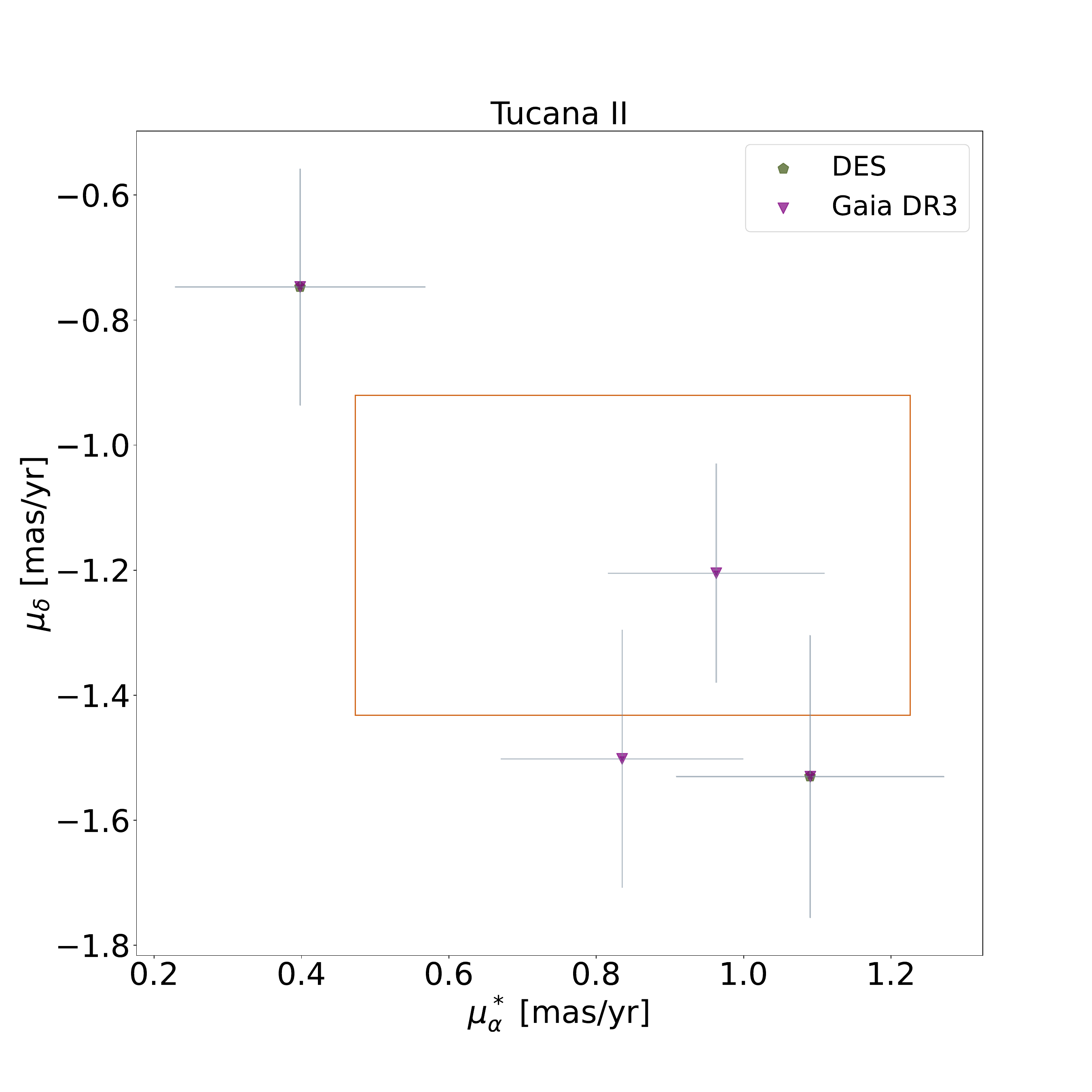}
    \caption{Proper motion boxes for four galaxies of our sample: Bootes I (\textit{upper left panel}), Ursa Major I (\textit{upper right panel}), DELVE 2 (\textit{lower left panel}) and Tucana II (\textit{lower right panel}). The limits for the boxes of Bootes I, Ursa Major I and Tucana II were defined according to the spectroscopic members reported in \cite{simon18}, while the box of DELVE 2 uses the galaxy's $\mu^*_\alpha$ and $\mu_\delta$ values stated in Table \ref{table:dataUFDs} obtained from \cite{Cerny2021a}. \label{fig:PMboxes}}
    \end{centering}
\end{figure*}

In order to compare the proper motion values of the RRL to those of the UFDs, we first defined a box for each galaxy inside which the proper motion values of the RRL candidates should fall to be considered a galaxy member. Example of these boxes can be seen in Fig. \ref{fig:PMboxes}. The size of the boxes were based on confirmed members for each galaxy available in the literature \citep{simon18, pace19, carlin18, longeard21, martinez21b, Cerny2022, garofalo21, fabrizio14, simon20, Jenkins2021, torrealba16b, Kirby2015}. The majority of these confirmed members were based on radial velocities, and/or proper motions from Gaia DR2 \citep{gaia18}. Here we updated the proper motions of all member stars with Gaia DR3 and eliminated some stars that were clearly outliers probably due to large differences in their proper motion in comparison to those previously reported in Gaia DR2. In a few cases, there are no list of members available in the literature. In those cases, we used the reported galaxy's mean proper motion as follows: for DELVE 2 and Virgo I we used the values provided in Table \ref{table:dataUFDs}, which were obtained from \cite{Cerny2021a} and \cite{mcconnachie2020}, respectively. Note that the proper motion error values stated in Table \ref{table:dataUFDs} for Virgo I is an average between those present in \cite{mcconnachie2020}; for the proper motion box of Eridanus IV, we used the limits presented in Figure 2 of \cite{Cerny2021b}; lastly, for the proper motion boxes of Leo T \citep{clementini12}, Pegasus III \citep{garofalo21} and Canes Venatici II \citep{Vargas2013} we used the galaxies' $\mu^*_\alpha$ and $\mu_\delta$ values as given in those references (which are the same as in Table \ref{table:dataUFDs}) without updating the values to Gaia DR3, since these three galaxies are beyond the Gaia limits. 

Once we had defined the proper motion boxes for each galaxy, we selected those RRL candidates with proper motion values that would fit in these boxes when considering their $1\sigma$ errors. 

This procedure was only followed for RRL from Gaia DR3, ZTF and PS1. We did not include the proper motion criterion in the selection of RRL candidates from DES data because this survey reaches fainter magnitudes than the Gaia limit. Therefore, the faintest RRL in DES do not have proper motions available. 

\section{Results} \label{sec:results}

\subsection{Final RRL sample} \label{subsec:foundRRL}

Out of the $45$ galaxies analyzed in this work, we find that $21$ of them have RRL stars according to our selection criteria stated in Sec. \ref{sec:methodology}. The remaining $24$ galaxies in which no RRL were found can be classified in two groups: those UFDs that truly do not have RRL and those that may have but the stars are not found in this work due to the limitations of the catalogs used. This is further dicussed in Sec. \ref{subsec:UFDnoRRL}. 

Our final RRL sample consists of $120$ stars which are listed in Table \ref{table:RRLdata}. The information provided for each star in Table \ref{table:RRLdata} is the following: (1) Gaia ID; (2) RA [deg]; (3) DEC [deg]; (4) G$_{Gaia}$ [mag]; (5-6) Gaia proper motion values; (7) RRL classification; (8) DES ID; (9) ZTF ID; (10) PS1 ID; (11-14) periods according to DES, ZTF, PS1 and Gaia respectively; (15) angular separation between the RRL and the host UFD, in $R_h$ units; and (16) name of the host UFD galaxy. The complete list of RRL can be found in the online version of this work. 

\begin{figure*}[!ht]
    \begin{centering}
%    \figurenum{3}
    \includegraphics[width=\columnwidth]{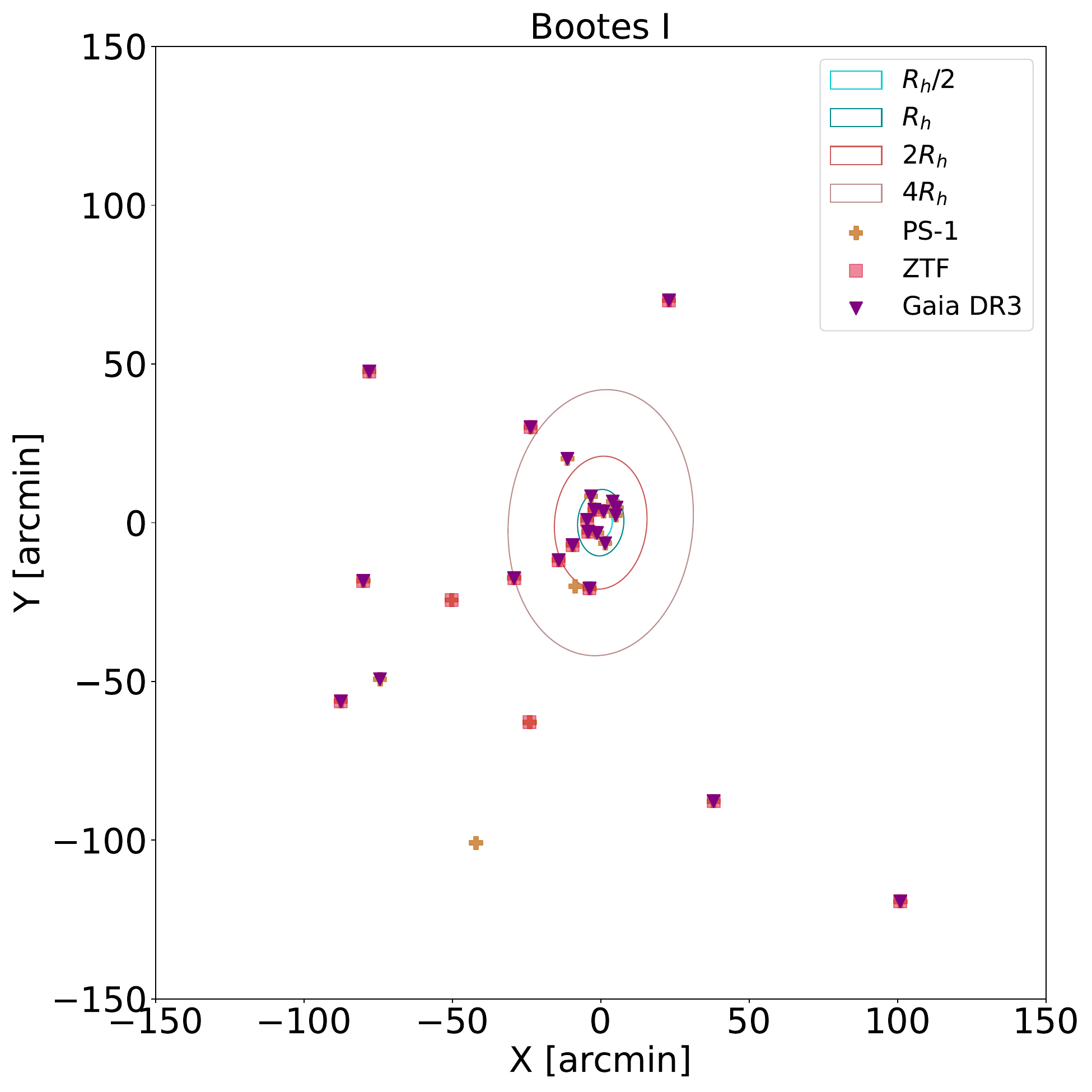}
    \includegraphics[width=\columnwidth]{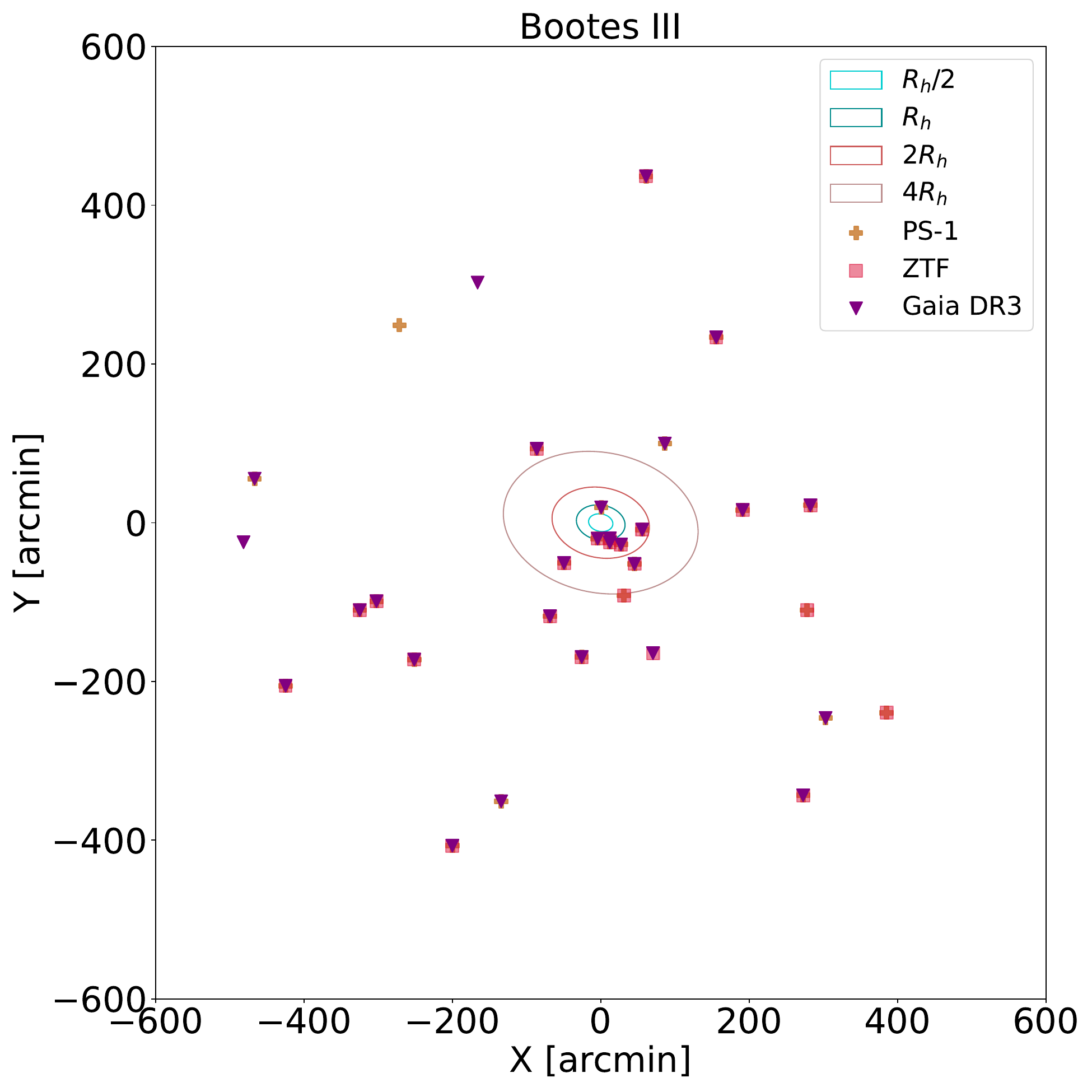}
    \includegraphics[width=\columnwidth]{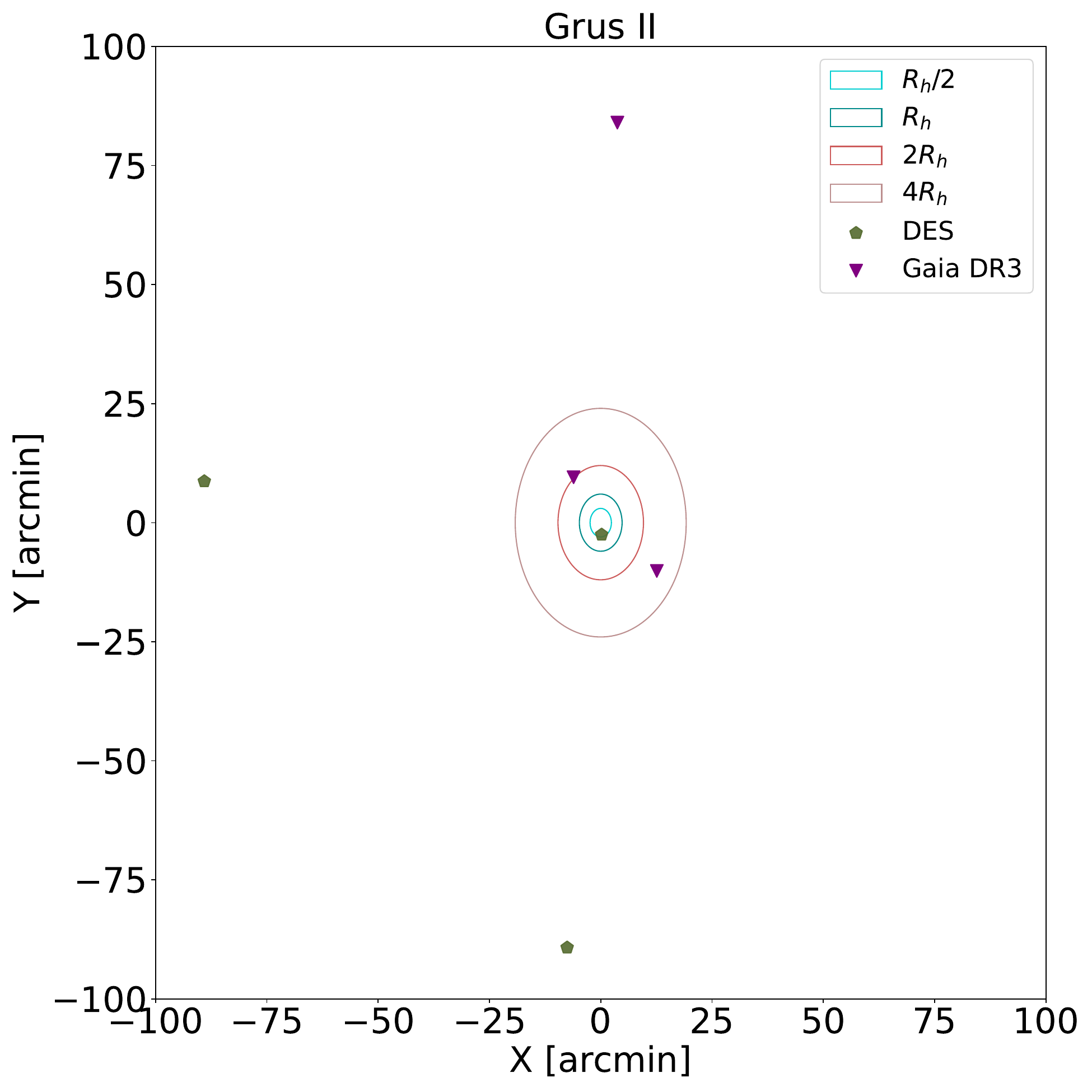}
    \includegraphics[width=\columnwidth]{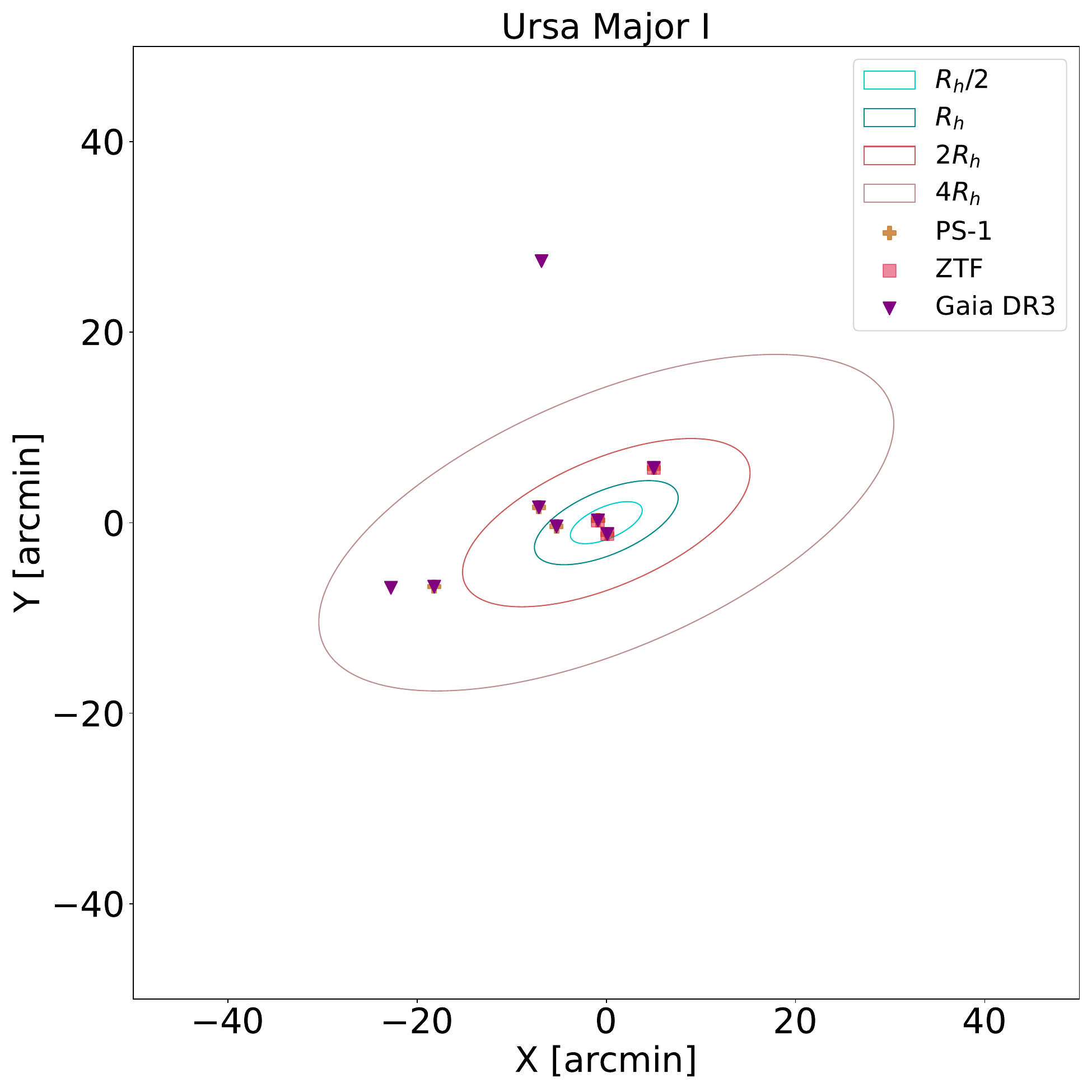}
    \caption{Spatial projection of the RRL found in four galaxies of our sample in planar coordinates centered on each UFD. Different symbols represent the catalog they are selected from. Four ellipses are plotted to represent the galaxy's size  ($R_h/2$, $R_h$, $2\, R_h$ and $4\, R_h$). We note that RRL are located at separations greater than $4\, R_h$ in all four cases. Plots for the remaining 17 UFDs in which we found RRL are available in the online version of this paper.
    \label{fig:SpatialProjection}}
    \end{centering}
\end{figure*}

Fig. \ref{fig:SpatialProjection} shows four examples of the spatial distribution of the RRL in planar coordinates centered on each UFD. The different symbols represent the catalog from which the RRL was selected. We also plotted four ellipses that represent $R_h/2$, $R_h$, $2\, R_h$ and $4\, R_h$ from the center of each galaxy. The equivalent figures for the rest of the galaxies in which we found RRL are available in the online version of this paper.

\begin{figure}
\centering
\includegraphics[width=0.47\textwidth]{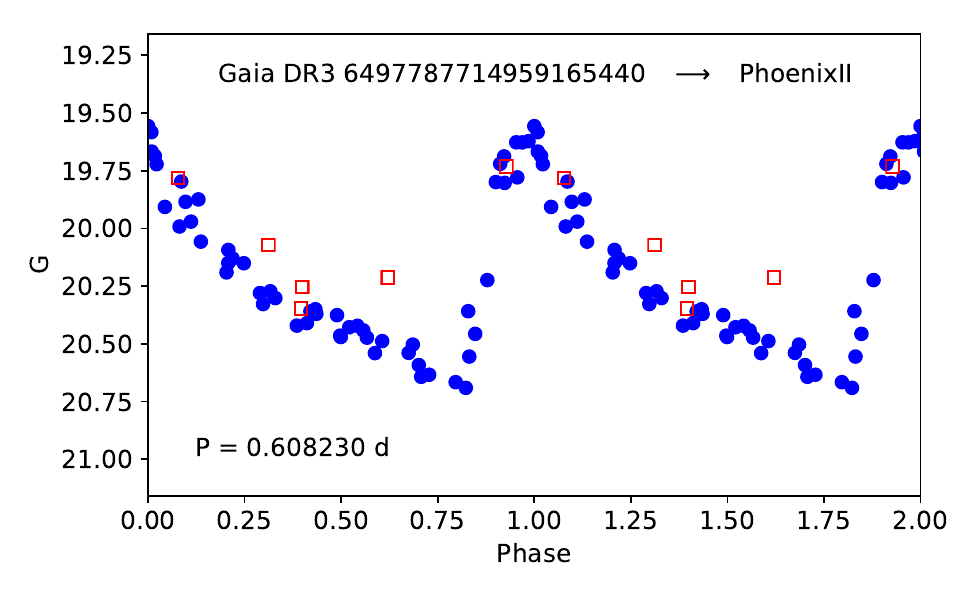}
\includegraphics[width=0.46\textwidth]{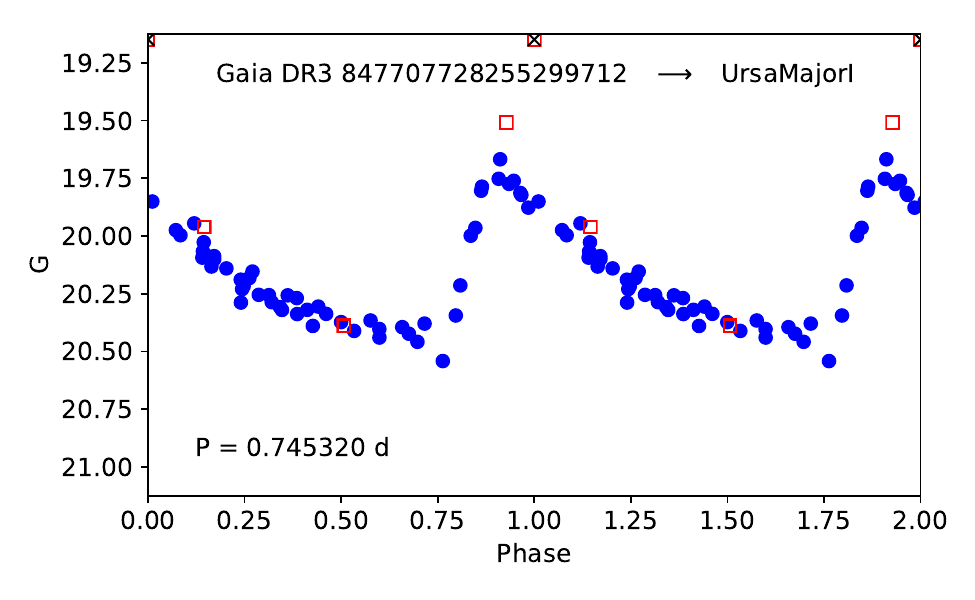}
\caption{Examples of Gaia DR3 lightcurves of two of the faintest RRL found in this work (in the Phoenix II and Ursa Major I galaxies).  The red squares and black x symbols mark measurements flagged in Gaia DR3 as noisy and rejected, respectively.}
\label{fig:lightcurve}
\end{figure}

Examples of lightcurves for two of the faintest RRL found in Gaia DR3 are shown in Fig.~\ref{fig:lightcurve}. The quality of the lightcurves is high even at these faint magnitudes, reassuring the selection of RRL is clean.

\begin{deluxetable*}{lrrrrrlrrlrrrrrl}
\rotate
\tabletypesize{\tiny}
\tablecolumns{16}
\tablewidth{0pc}
\tablecaption{Final sample of RRL found in $21$ different UFDs. \label{table:RRLdata}}
\tablehead{
ID$_{Gaia}$ &    RA &   DEC &  $G_{Gaia}$ &  $\mu_{\alpha} \cos(\delta) (Gaia)$ &  $\mu_{\delta} (Gaia)$ & Type$^\dag$ &  ID$_{DES}$ & ID$_{ZTF}$ & ID$_{PS1}$ & $P_{DES}$ &  $P_{ZTF}$ &  $P_{PS1}$ &  $P_{Gaia}$ &    Sep/R$_h$ &       Galaxy \\
                  & [deg] & [deg] & [mag] & [mas/yr] & [mas/yr] & & & & & [days] & [days] & [days] & [days] & & \\
                  }
\startdata
4918034941751885696 &   1.07229 & -59.37689 &      17.54 &  -0.053 & -1.5349 & RRab & 1041991456 &                   - &      - &  0.57282 &        - &        - &  0.57282 &   14.03 &     TucanaIII \\
4917942612838673024 &   0.99559 & -60.06946 &      19.40 &  0.8997 & -1.7487 &  RRc &          - &                   - &      - &        - &        - &        - &  0.36598 &    5.32 &      TucanaIV \\
6487761169641710464 & 357.54223 & -62.06401 &      19.30 &  0.8134 & -0.7239 &  RRc &          - &                   - &      - &        - &        - &        - &  0.29653 &   14.09 &      TucanaIV \\
                  - &   2.36329 & -60.98564 &      19.23 &       - &       - & RRab & 1045709754 &                   - &      - &  0.62125 &        - &        - &        - &    8.43 &      TucanaIV \\
6567699405896736384 & 331.10818 & -45.04026 &      19.13 &    0.38 & -0.6487 &  RRc &          - &                   - &      - &        - &        - &        - &  0.36131 &   14.02 &        GrusII \\
6561357319748782976 & 331.32574 & -46.60868 &      18.74 &  0.4474 & -1.9714 & RRab &          - &                   - &      - &        - &        - &        - &  0.62057 &    3.12 &        GrusII \\
6561426485901552640 & 330.87292 & -46.28102 &      19.11 &   0.496 & -1.5272 &  RRc &          - &                   - &      - &        - &        - &        - &   0.4044 &    2.03 &        GrusII \\
                  - & 328.87317 & -46.27519 &      18.61 &       - &       - & RRab &  940908371 &                   - &      - &  0.57105 &        - &        - &        - &   18.61 &        GrusII \\
                  - & 331.02493 & -46.48205 &      18.64 &       - &       - & RRab &  947555289 &                   - &      - &  0.41952 &        - &        - &        - &    0.42 &        GrusII \\
                  - & 330.83239 & -47.92666 &      18.55 &       - &       - & RRab &  947930606 &                   - &      - &  0.60425 &        - &        - &        - &   14.95 &        GrusII \\
6864048408410304896 & 298.15855 & -22.05848 &      19.76 &  0.3723 &   -0.67 &  RRc &          - &                   - &      - &        - &        - &        - &  0.30785 &    0.34 & SagittariusII \\
6864423994704275200 & 298.14951 & -22.03330 &      19.68 &  0.0635 & -0.5289 &  RRc &          - &                   - &      - &        - &        - &        - &  0.31856 &    1.25 & SagittariusII \\
6864422757758521984 & 298.23647 & -22.06901 &      19.64 & -0.6756 & -0.9773 &  RRc &          - &                   - &      - &        - &        - &        - &  0.40651 &    2.30 & SagittariusII \\
6864422993976659968 & 298.18499 & -22.05134 &      19.59 & -0.7179 & -0.9086 & RRab &          - & 6864422993976659968 &      - &        - &  0.66549 &        - &  0.66565 &    0.78 & SagittariusII \\
                  - & 298.05471 & -21.71638 &      19.44 &       - &       - & RRab &          - & 6865195302117134336 &      - &        - &  0.54079 &        - &        - &   12.84 & SagittariusII \\
  87207528534363264 &  34.75026 &  20.10978 &      18.55 &  1.4086 & -0.3918 & RRab &          - &   87207528534363264 &      - &        - &  0.74948 &        - &  0.74949 &    1.74 &       SegueII \\
                  - & 344.19906 & -50.18682 &      20.75 &       - &       - & RRab &  983283631 &                   - &      - &  0.63549 &        - &        - &        - &    1.22 &         GrusI \\
1243152071642576896 & 208.67327 &  15.30334 &      19.31 & -1.4792 & -0.0398 & RRab &          - & 1243152071642576896 &     ** &        - &  0.62692 &  0.62694 &  0.62691 &   11.26 &       BootesI \\
1226522469372990208 & 211.74214 &  12.52060 &      19.73 &  0.0338 & -1.0351 & RRab &          - & 1226522469372990208 &     ** &        - &  0.64768 &  0.64766 &  0.64766 &   17.80 &       BootesI \\
1229801659723433600 & 210.67042 &  13.05105 &      19.14 &  -1.331 & -0.3902 & RRab &          - & 1229801659723433472 &     ** &        - &  0.59843 &  0.59845 &  0.59841 &    9.97 &       BootesI \\
1230741020611316224 & 209.51825 &  14.22200 &      19.44 & -0.5014 & -0.4917 & RRab &          - & 1230741020611316224 &     ** &        - &  0.66164 &   0.6617 &  0.66168 &    4.00 &       BootesI \\
1230772597210842624 & 209.95568 &  14.16816 &      19.43 &  0.1171 &  -1.277 &  RRc &          - & 1230772597210842624 &     ** &        - &  0.40118 &  0.40118 &  0.40116 &    2.01 &       BootesI \\
\enddata
\tablecomments{ ** Star was found in PS1 but no IDs were not provided in \cite{sesar17}'s catalog.}
\tablecomments{ $\dag$ \textit{DES}: all of the RRL reported in this catalog \citep{Stringer2021} are type RRab; \textit{ZTF}: no type is reported in the RRL catalog by \cite{Huang2022}, but the authors suggest a classification based on the stars' periods: type RRab if its greater than $0.45$ days and type RRc if its lower than $0.45$ days; \textit{PS1}: \cite{sesar17} propose a classification based on the ``\textit{score$_3$}" parameter of the stars in their catalog: type RRab if $s_{3,ab} > 0.8$ and type RRc if $s_{3,c} > 0.55$.}
\tablecomments{ Only a portion of this table is shown here to demonstrate its form and content. The complete list of RRL is available in the online version of this paper.}
\end{deluxetable*}

\subsection{Possible Contamination by the Halo and/or Halo Substructures}
\label{subsec:contamination}

As discussed before, in general we expect negligible contamination in the selected RRL around each UFD from the halo field population. However, there are some special cases in which contamination may be important. Firstly, although RRL are rare in the distant halo, if the search area is very large and/or the UFD is not at a large distance from the Galactic center, we may expect random halo RRL field stars within the selected sample. Most of the UFDs are very small and consequently the search area is also small. However, Bootes III is exceptionally large and halo contamination must be addressed. Secondly, some galaxies are close enough to other stellar systems or halo substructures, such as the Small Magellanic Cloud (SMC) and stellar streams, that may also have contamination by RRL from those substructures. We discuss these special cases in the following subsections.

\subsubsection{Galaxies Near the Sagittarius Stream: Bootes I, Bootes II, and Bootes III} \label{subsubsec:cont_sgrstream}

\begin{figure}[!h]
\plotone{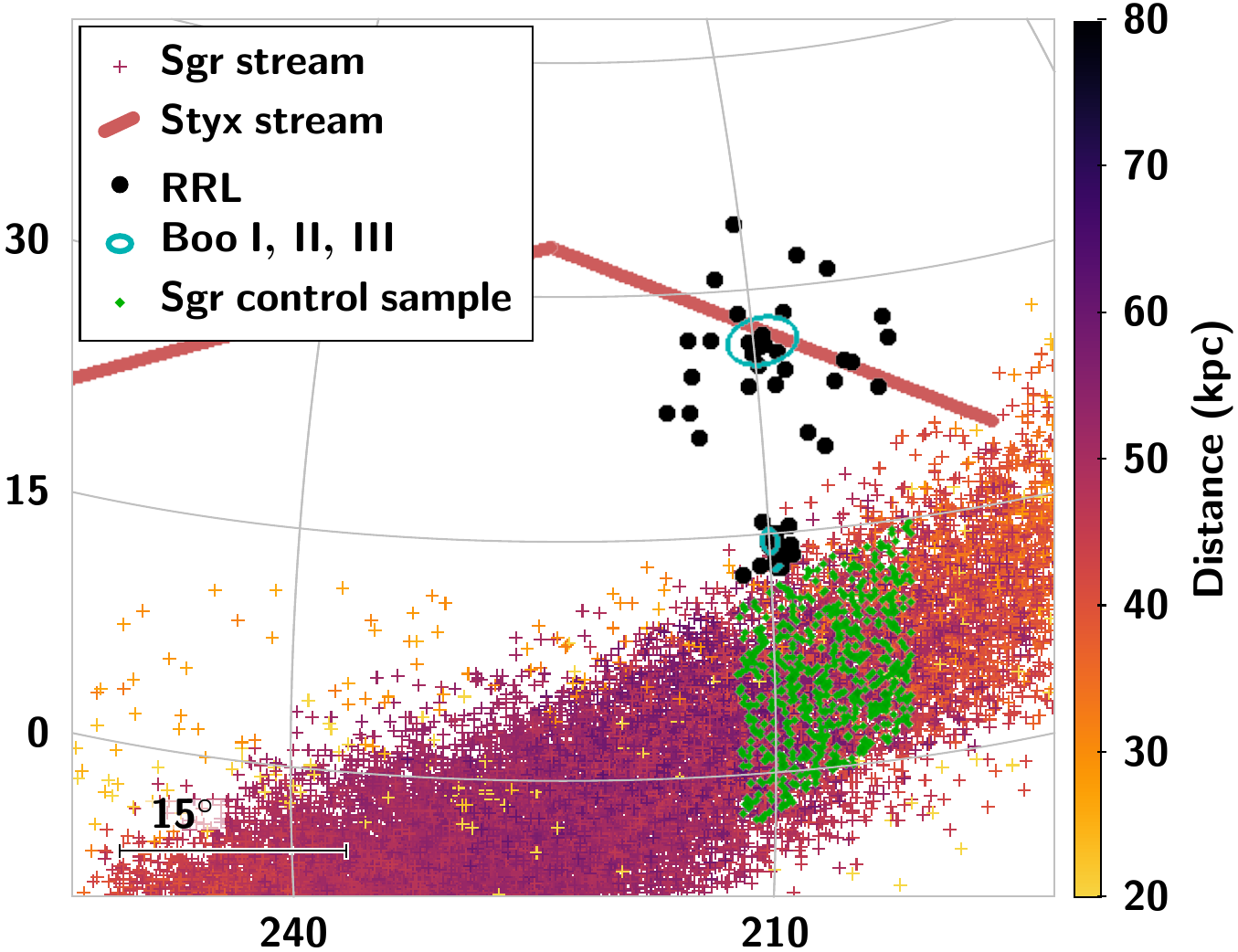}
\caption{Equatorial coordinates (x,y=RA,DEC) map of the RRL stars (black circles) selected as members of the Bootes I, II, III galaxy. The blue ellipses encloses $4\, R_h$ of each galaxy. Bootes III is the largest one at $\delta=26\fdg 8$, while Bootes II is the smallest and further South one. The model of the Sgr stream by \citet{vasiliev21} is shown with $+$ symbols whose color scales with the heliocentric distance. Green $+$ symbols are Gaia DR3 RRL in the Sgr stream selected as a control sample (see text). The solid line approximately traces the Styx stream \citep{grillmair09}.} 
\label{fig:styx}
\end{figure}

Bootes III is a large UFD, which is very likely in the process of being disrupted \citep{carlin09,carlin18}. It was discovered by \citet{grillmair09} as a strong overdensity in SDSS data. The same paper reports a stellar stream, Styx, which passes through Bootes III and has a similar heliocentric distance (Fig.~\ref{fig:styx}), making a relationship between both structures very likely. Indications that the galaxy is being disrupted are its large size \citep[$R_h = 33\farcm 03$,][]{moskowitz20}, large velocity dispersion, and an orbit that brings Bootes III as close as 7.8 - 12 kpc from the Galactic center at pericenter \citep{carlin18,pace22}. 

\begin{figure}
\centering
\includegraphics[width=0.47\textwidth]{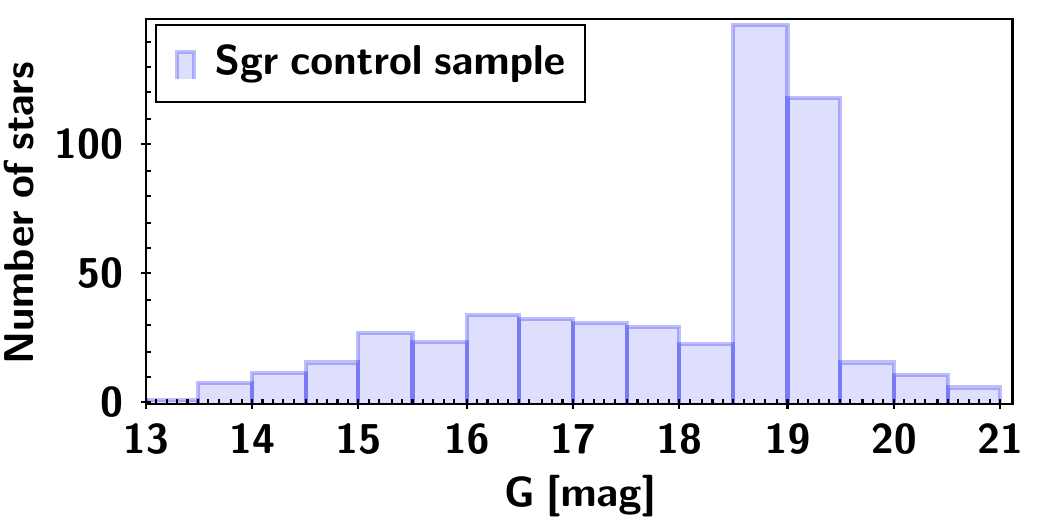}
\includegraphics[width=0.46\textwidth]{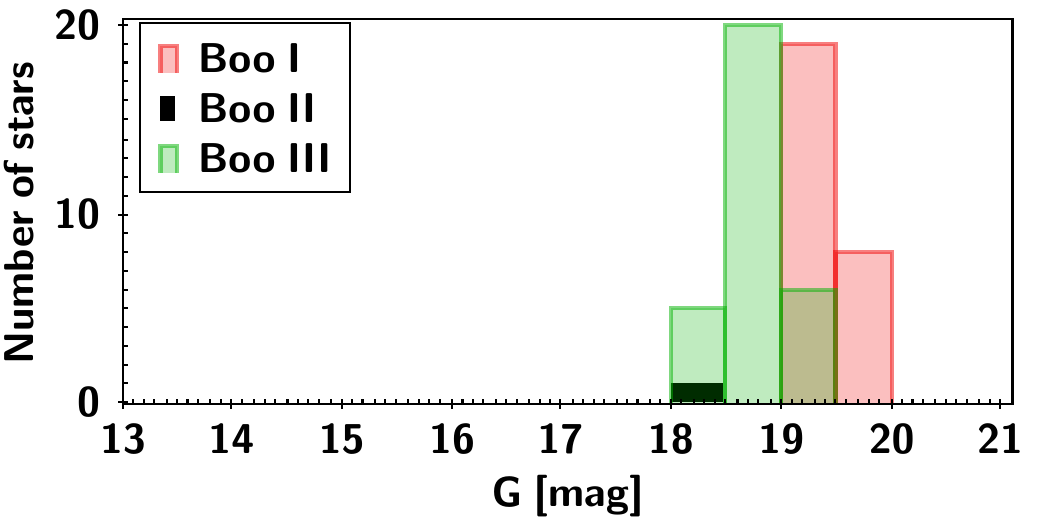}
\includegraphics[width=0.47\textwidth]{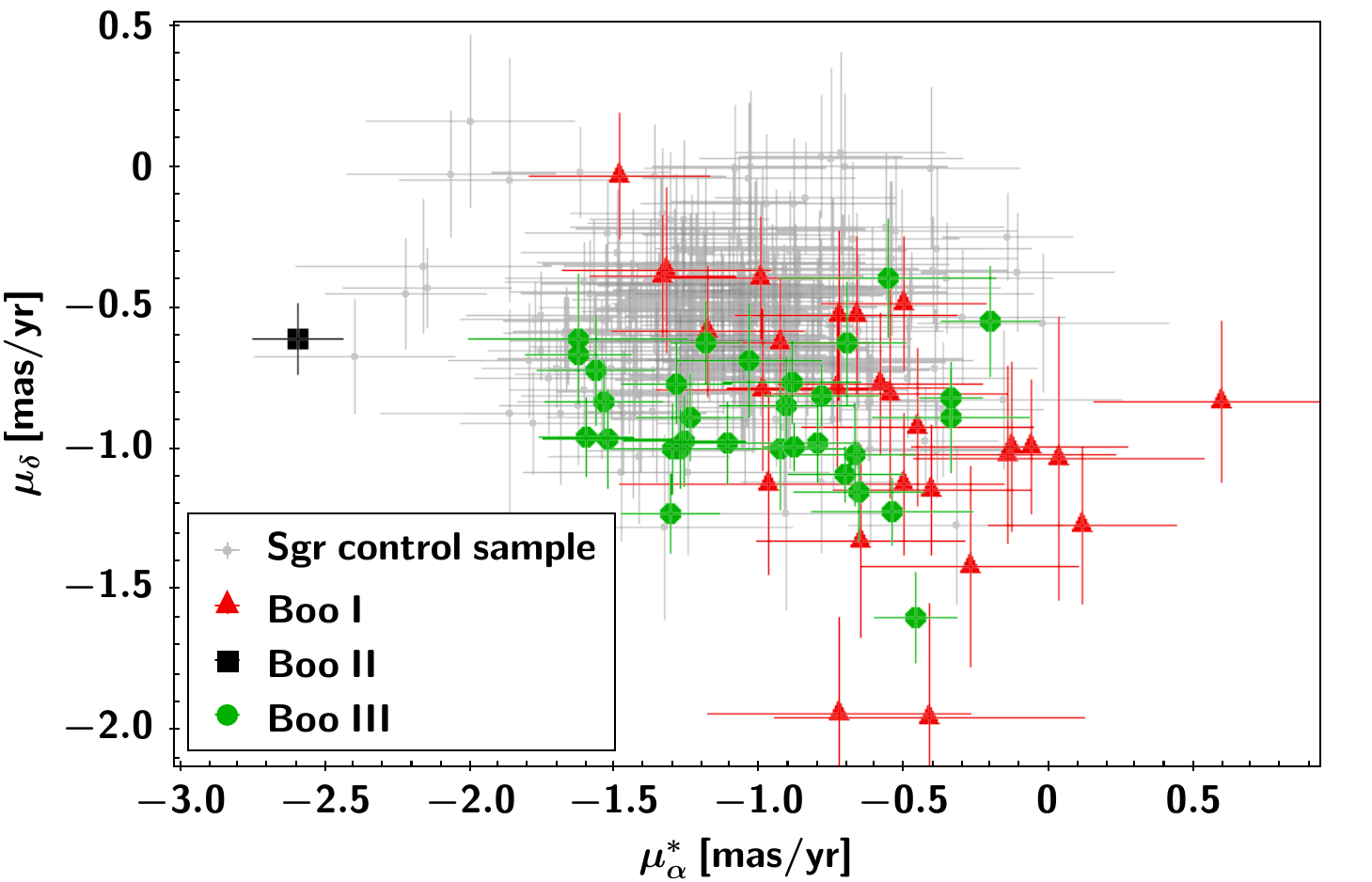}
\caption{\textit{Top panel}: distribution of G magnitude of RRL selected from Gaia DR3 in the Sgr stream (green $+$ symbols in Fig.~\ref{fig:styx}). \textit{Middle panel}: distribution of G magnitudes of RRL selected as possible members of the Bootes I (red), Bootes II (black), and Bootes III (green) galaxies. \textit{Bottom panel}: Gaia DR3 proper motions of the RRL in Bootes I, II and II compared with the control sample in the Sgr stream.}
\label{fig:Sgr}
\end{figure}

\citet{vivas20} identified seven RRL in Bootes III based on Gaia DR2 in an area of $2\degr$ around the galaxy, or about $4 R_h$. Here we explored a much larger area around the galaxy, finding $32$ RRL likely associated with Bootes III since they share the same magnitude (as can be seen in the upper right panel of Fig. \ref{fig:GvsBPRP}) and proper motions as the galaxy. They spread however in an area as large as the explored region, with quite a uniform distribution (as shown in the upper right panel of Fig. \ref{fig:SpatialProjection}). In particular, the selected RRL stars do not seem to trace the Styx stream, suggesting Styx is not responsible for many of these RRL. Although in principle all these RRL stars could be indeed debris material from Bootes III and/or Styx, we noticed the Sagittarius (Sgr) stream gets as close as $\sim10 \degr$ to the South of Bootes III \citep{vasiliev21}. Coincidentally, in this part of the sky, the Sgr stream has a distance of $\sim 35-45$ kpc, not too different than the distance to Bootes III (Fig.~\ref{fig:styx}). Thus, some contamination by Sgr stream stars may exist among the candidates to debris from Bootes III. The fact that there seems to be more RRL stars toward the South of Bootes III compared with the Northern side, indicates there may be indeed Sgr contamination in our selection. This can be seen in Fig.~\ref{fig:Sgr} which shows that there is overlap between both the G magnitudes and the proper motions of the selected Bootes III RRL and the ones in the Sgr stream (control sample, green $+$ symbols in Fig.~\ref{fig:styx}). The RRL in Bootes III are slightly brighter ($\langle G \rangle = 18.76$) than the bulk of the Sgr stream stars ($\langle G \rangle = 18.98$), but there is still enough overlap to believe some degree of contamination is possible.

On the other hand, since Bootes III is relatively close to the Sun ($47$ kpc) and the search area is very large in this case ($\sim 190$ sq deg), we expect contamination by random halo stars. By integrating the RRL number density radial profile \citet{medina18} between $37$ and $47$ kpc (the range of distance of the stars found), we estimate as many as $16$ RRL in the searched area around Bootes III. Although it is expected that the proper motion cut we implemented can get rid of some of these Halo contaminants, no clean separation is seen. Indeed, there were $45$ RRL selected before the proper motion cut. If $\sim 16$ are halo stars, we expect the excess of stars due to Bootes III to be $\sim 29$ RRL, confirming that we expect some small degree of contamination among the $32$ RRL finally selected. Radial velocities would be needed to confirm the association of the RRL with either Bootes III, Sgr stream or the Halo. Nevertheless, we expect several of the RRL to be true members of Bootes III, with some of them extending to large distances from its center.

\bigskip

Bootes I has a large population of RRL with 15 known from \cite{siegel06} and \cite{dallora06}, and an additional one found in Gaia DR2 by \citet{vivas20}. Two of those previously known stars seem to be extra-tidal stars. Our search found a total of $27$ RRL around Bootes I, including the $16$ mentioned RRL already reported. The new stars range in distance between $2.47$ and $17.80\, R_h$ from the center of the galaxy. We notice however that Bootes I is just on the edge of the Sgr stream (Fig.~\ref{fig:styx}) and, similar to the case of Bootes III described above, the proper motions and G magnitudes of the RRL stars overlap (Fig.~\ref{fig:Sgr}). The range of G magnitudes of the Bootes I RRL range from $19.14$ mag to $19.73$ mag, slightly fainter than the range of Sgr stars ($G = 18.5 - 19.5$ mag). With the current data we are unable to isolate the true members but we suspect that in this case, most of the stars in the outskirts of Bootes I are indeed Sgr stream stars. It is possible that the 4 new RRL on the northern side of Bootes I (to the other side of Sgr stream) may be true members of the galaxy. The spatial distribution of the RRL for this (and every other) galaxy are provided in the electronic form on this paper (Fig.~\ref{fig:SpatialProjection}).

\bigskip

Finally, Bootes II is embedded in the Sgr stream (Fig.~\ref{fig:styx}). We found only one RRL associated with this star, which was previously known \citep{sesar14}. The star is located well within $1\, R_h$ from the center of Bootes II. This RRL is brighter than the bulk of RRL in the Sgr stream and the proper motion also separates well from Sgr stars (Fig.~\ref{fig:Sgr}). Thus, in this case, we are confident to believe this RRL is indeed a member of the UFD.

\subsubsection{Galaxies near the SMC: DELVE 2, Tucana IV, and Tucana III} \label{subsubsec:cont_SMC}

\begin{figure}[!h]
\plotone{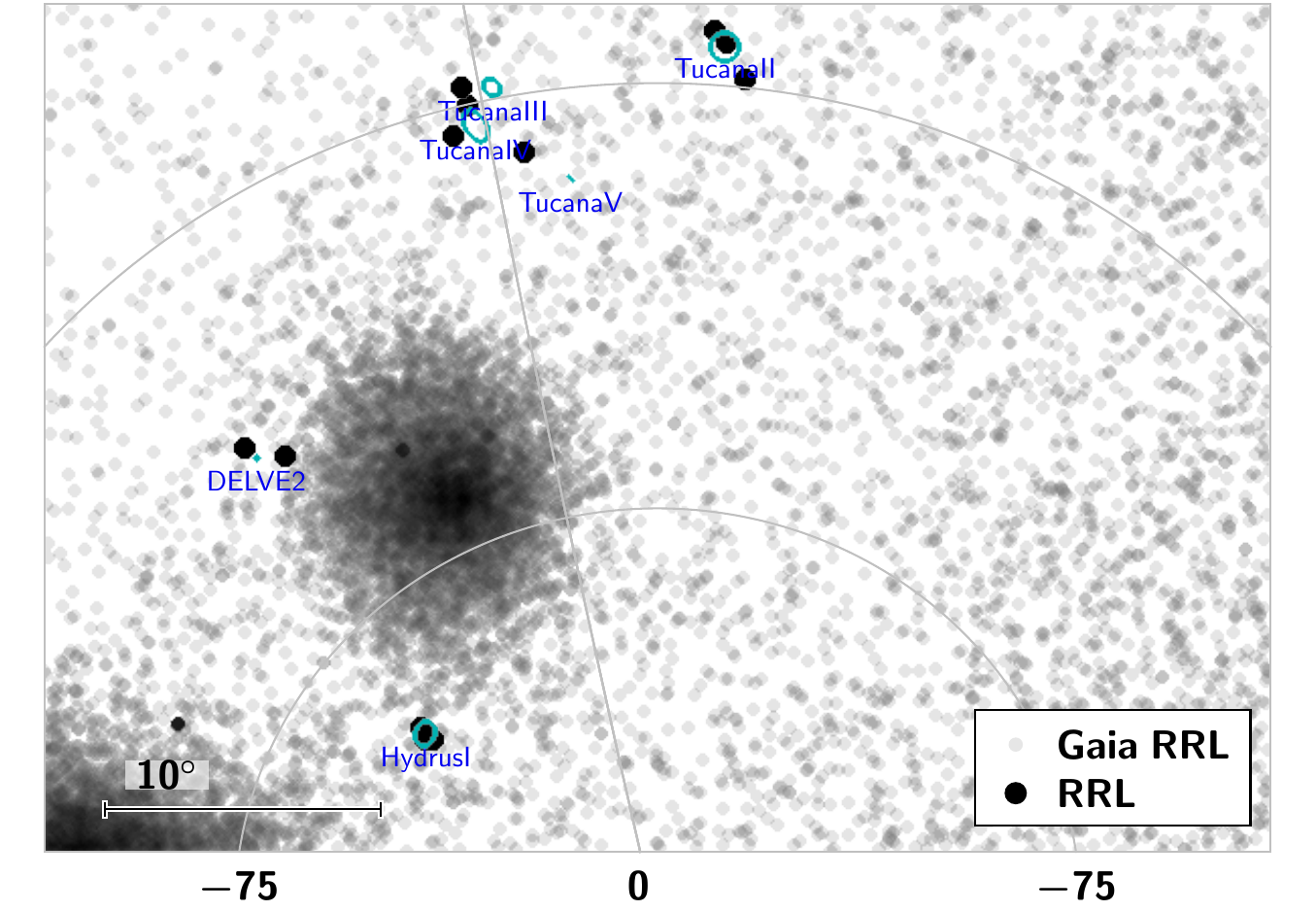}
\caption{Map in equatorial coordinates of UFDs in the neighborhood of the SMC. The grey background shows all RRL in Gaia DR3 in the neighborhood, with the large concentration of stars being the SMC.  Black circles show the RRL stars selected as members of the Tucana II, Tucana III, Tucana IV, Hydrus I, and DELVE 2 galaxies. We found no RRL associated with Tucana V. The blue ellipses encloses $4\, R_h$ of each galaxy, which is very small in the case of DELVE 2 ($R_h = 1\farcm02$), and Tucana V ($R_h= 1\farcm0$).} 
\label{fig:nearSMC}
\end{figure}

Several UFDs are located in the neighborhood of the SMC as seen in Fig.~\ref{fig:nearSMC}. The SMC is known to have a very extended population \citep{Massana2020} and can contaminate the sample of RRL selected in those UFDs. Of the 5 nearby galaxies, we found no RRL associated with Tucana V. On the other hand, we discard contamination from the SMC in Hydrus I and Tucana III because the distances to those systems are very different to that of the SMC (see Table~\ref{table:dataUFDs}). 

\begin{figure}
%\centering
\raggedleft
\includegraphics[width=0.47\textwidth]{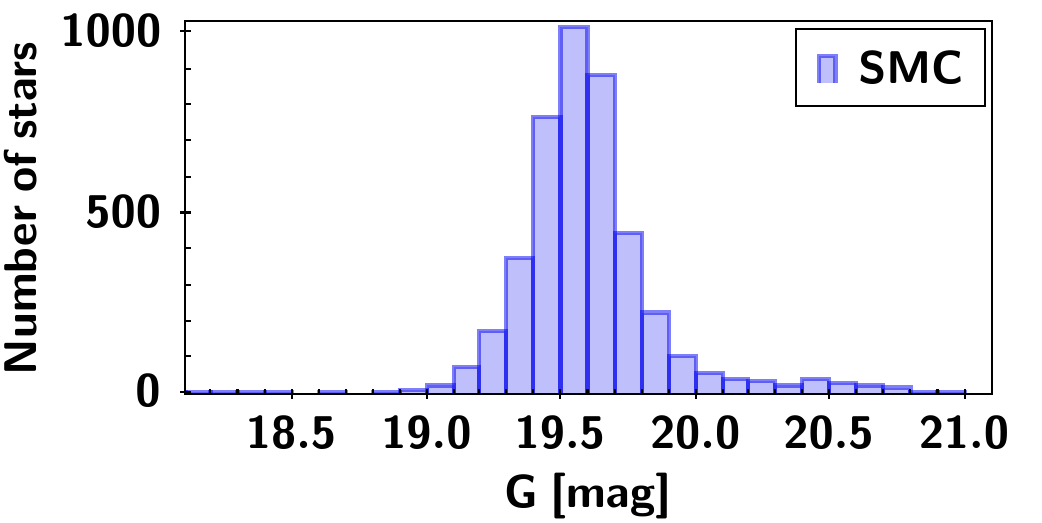}
\includegraphics[width=0.46\textwidth]{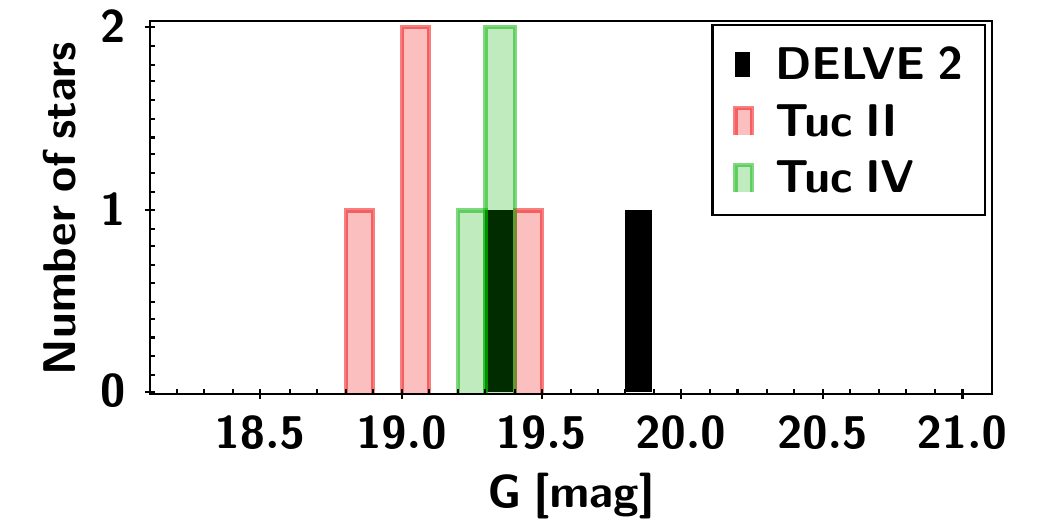}
\includegraphics[width=0.47\textwidth]{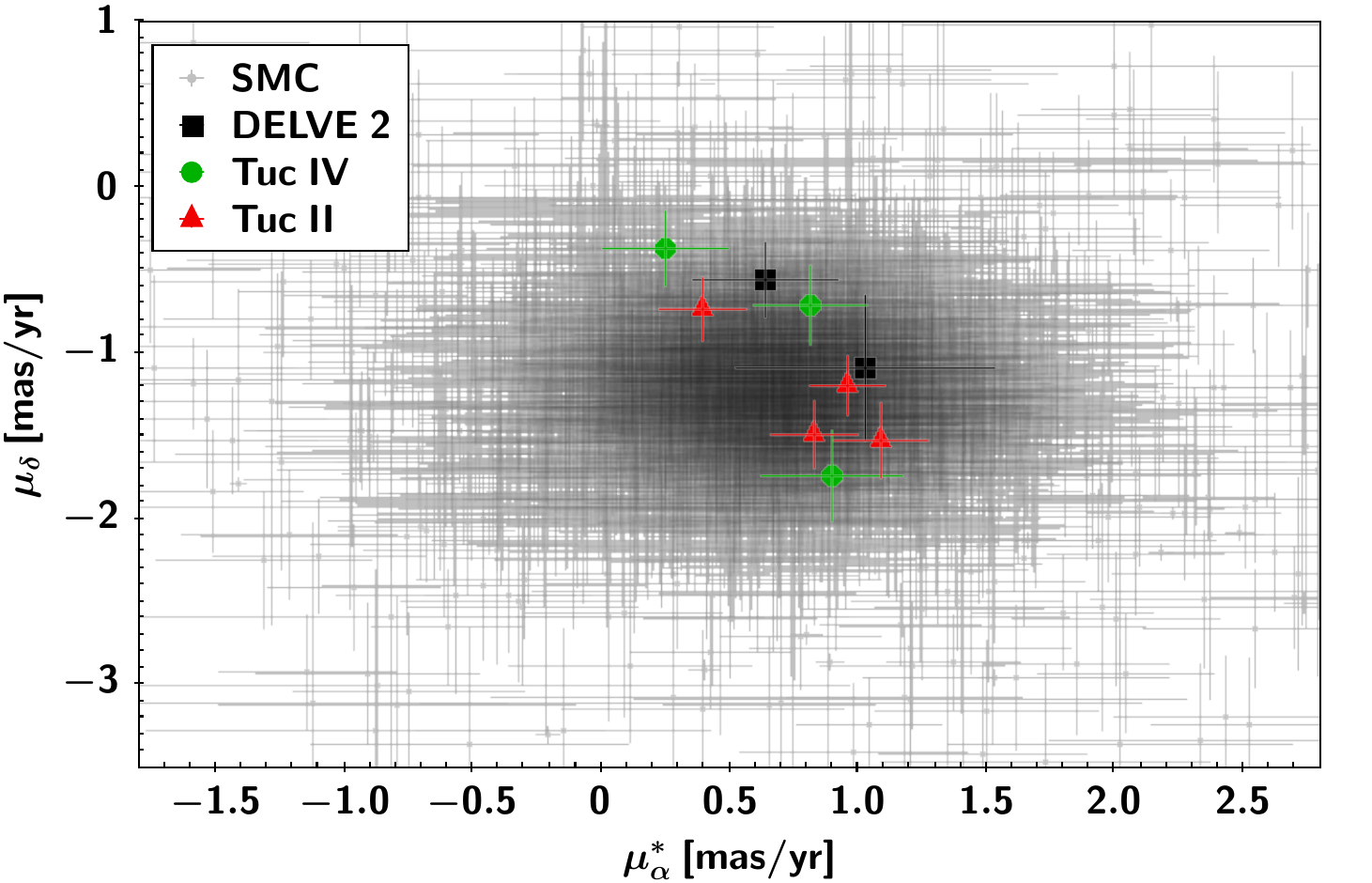}
\caption{\textit{Top panel}: distribution of G magnitude of RRL selected from Gaia DR3 in the SMC. \textit{Middle panel}: distribution of G magnitudes of RRL selected as possible members of the Tucana II (red), DELVE 2 (black), and Tucana IV (green) galaxies. \textit{Bottom panel}: Gaia DR3 proper motions of the RRL in DELVE 2, Tucana II and Tucana IV compared with the proper motions of RRL in the SMC.}
\label{fig:SMC}
\end{figure}

In the upper and middle panels of Fig.~\ref{fig:SMC} we show the distribution of G magnitudes of RRL in the SMC as well as in DELVE 2, Tucana II, Tucana IV. The lower panel shows the distribution of proper motions. These plots show that there is enough overlap in magnitude (i.e. distance) and proper motions to consider possible contamination from the SMC in those 3 UFDs.

DELVE 2 is located at only $6\fdg 9$ from the SMC. Although the two RRL stars found in our search are within $1\degr$ from the UFD, the galaxy is tiny and they are indeed $>35\; R_h$ from the center of DELVE 2. In addition, the magnitudes and proper motions of the two RRL completely overlap with those of the SMC (Fig.~\ref{fig:SMC}). Thus, it is clear in this case that those stars are most likely members of the SMC rather than of DELVE 2, in agreement with the conclusions by \citet{Cerny2021a}.

Tucana IV may also suffer contamination from the SMC. All the three RRL we found in this galaxy are far away from its center, at 5, 8, and 14 $R_h$. Those stars also have magnitudes and proper motions overlapping with the SMC (Fig.~\ref{fig:SMC}), which is at an angular distance of $12\fdg 9$ from the UFD. It is very possible that these three stars are also SMC distant members.

In the case of Tucana II, a possible contamination from the SMC is not that clear. Not only the galaxy is farther away from the SMC, at $18\fdg 5$, but out of its 4 RRL, 2 of them are located within $1.1 R_h$. The most external ones (at 5 and 12 $R_h$) are also the brightest ones in the group ($G = 19.07, 18.97$ mag), which may be too bright for the SMC. Although confirmation via radial velocities is needed, we believe that the scenario of contamination from SMC is weak in the case of Tucana II.

\subsubsection{Grus II and the Chenab/Orphan Stream}

\citet{martinez19} searched for RRL in Grus II, finding 3 of them in the galaxy. However, further analysis made those authors realize that there may be two substructures overlapping in this part of the sky: Grus II and the Chenab/Orphan stream \citep{Koposov2019} that passes several kpc in projection in front of Grus II. Here we recovered the 3 known stars, and found 3 additional ones, which are located outside the area searched by \citet{martinez19}.  As in \citet{martinez19} we also found a bimodal distribution in the magnitudes of the RRL (see lower left panel in Fig.~\ref{fig:GvsBPRP}), with two stars having a magnitude of $G=19.1$, while the other 4 are $\sim 0.5$ mag brighter. The two fainter stars are hence associated with Grus II, while the other 4 are very likely contamination by the Chenab/Orphan stream. 

\subsection{UFDs with no RRL} \label{subsec:UFDnoRRL}

We did not find any RRL in $24$ UFDs of our sample. These galaxies are listed in Table \ref{table:UFDs_noRRL} and they are separated in two groups: those galaxies that truly do not seem to have RRL, and those in which we do not find any but could nevertheless have some RRL that are outside the limits of the surveys used in this work.

\begin{deluxetable}{lcc}
\tabletypesize{\normalsize}
\tablecolumns{3}
\tablewidth{0pc}
\tablecaption{UFDs of our sample with no RRL according to our selecion criteria, sorted by their distance. \label{table:UFDs_noRRL}}
\tablehead{
Galaxy &  Distance &  Expected $m_G$ \\
                  &  [kpc]    &  [mag] \\ 
}
    \startdata
    \sidehead{\textbf{UFDs with no RRL}} \\
        Draco II &     21.5 &      17.13 \\
       Carina III &     27.8 &      17.69 \\
    Triangulum II &     28.4 &      17.74 \\
         Cetus II &     30.0 &      17.86 \\
     Reticulum II &     30.0 &      17.86 \\
        Willman I &     38.0 &      18.37 \\
         Tucana V &     55.0 &      19.17 \\
      Eridanus IV &     76.7 &      19.89 \\
    Horologium II &     78.0 &      19.93 \\
     Horologium I &     79.0 &      19.96 \\
          Virgo I &     87.0 &      20.17 \\
     Eridanus III &     87.0 &      20.17 \\
    Reticulum III &     92.0 &      20.29 \\
      Aquarius II &    107.9 &      20.64 \\
      \sidehead{\textbf{UFDs outside of our detection range}} \\
      Centaurus I$^*$ &    116.3 &      20.80 \\
         Hercules$^*$ &    133.0 &      21.09 \\
         Hydra II$^*$ &    151.0 &      21.37 \\
           Leo IV$^*$ &    154.0 &      21.41 \\
    Canes Venatici II$^*$ &    159.0 &      21.48 \\
          Leo V$^*$ &    173.0 &      21.66 \\
      Pegasus III$^*$ &    174.0 &      21.67 \\
        Pisces II$^*$ &    175.0 &      21.69 \\
        Columba I &    183.0 &      21.78 \\
        Leo T$^*$ &    409.0 &      23.53 \\
    \enddata
\tablecomments{The third column shows the expected magnitude that the RRL would have if they were located at distances similar to the UFDs. UFDs with known population of RRL that we did not recover are marked with an asterisk ($*$): Centaurus I \citep{martinez21b}, Hercules \citep{garling18}, Hydra II \citep{vivas16}, Leo IV \citep{moretti09}, Leo V \citep{medina18}, Canes Venatici II \citep{Greco2008}, Pegasus III $\&$ Pisces II \citep{garofalo21}, and Leo T \citep{clementini12}. }
\end{deluxetable}

The third column in Table \ref{table:UFDs_noRRL} shows the expected magnitude that the RRL would have if they were located at distances similar to that of the UFD. On one hand, for those UFDs which have expected magnitudes brighter than $G\sim 21$ mag, Gaia DR3 should have been able to detect them if there were any. Thus, we can confirm that in these cases the UFDs do not have indeed any RRL. In agreement with the analysis done by \cite{vivas20}, we do not find any RRL in Draco II, Carina III, Triangulum II, Cetus II, Willman I, Tucana V, Horologium II, Horologium I and Virgo I. Neither Eridanus IV nor Aquarius II were part of the sample studied by \citet{vivas20}. Regarding Eridanus III and Reticulum III, although \cite{vivas20} found one RRL in each galaxy using Gaia DR2, we do not find any in this work. The two stars from \citet{vivas20} do not even exist in the RRL catalog provided by Gaia DR3, which means that they were likely misclassified in the previous data release. 

On the other hand, for those galaxies with expected magnitudes for their RRL fainter than $G\sim 21$ mag, Gaia is just not able to detect them because they are too faint. Although DES reaches fainter magnitudes than Gaia, its sky coverage is limited and only a few UFDs fall within its footprint. The only one of these distant UFDs that is located within DES footprint is Columba I, but the expected magnitude for its RRL is very faint ($\sim 21.78$ mag). The DES RRL survey's completeness strongly declines at these faint magnitudes \citep{Stringer2021}, which may be the reason why we do not find any RRL in this galaxy. All of the other UFDs (Centaurus I, Hercules, Canes Venatici II, Leo V, Leo IV, Pegasus III, Pisces II and Leo T) are not detected by DES due to their positions in the sky. However, except Columba I, all of these distant galaxies have had dedicated searches for variable stars (see Table \ref{table:UFDs_noRRL} for references). A dedicated search for variable stars in Columba I is recommended. In the case of Centaurus I, the expected magnitude of the RRL is $\sim 20.80$ mag, within the Gaia's limits but approaching its limiting magnitude. Centaurus I is known to have RRL: \cite{martinez21b} report three of them based on their own multi-epoch \textit{giz} DECam observations of this UFD. We found none neither in Gaia nor in the other surveys. The completeness of Gaia decreases as it approached its faint limit and that is probably the reason why these stars were not detected in this work.

The distant UFDs with known RRL population have had dedicated searches of variable stars but the coverage is not necessarily large enough to asses the existence of extended populations. An exception is Hercules. The galaxy has been reported to have $12$ RRL by \cite{Musella2012} and \cite{garling18}, where the latter work showed that Hercules has indeed RRL outside its tidal radius. We compiled the RRL from these distant galaxies, as well as a few RRL in nearby galaxies that we did not recover (in Tucana III \citep[$5$ RRL,][]{vivas20}, Sagittarius II \citep[$1$ RRL,][]{joo19}, Grus I \citep[$1$ RRL,][]{martinez19} and Eridanus II \citep[62 RRL,][]{MartinezVazquez2021a}) in a separate list labeled \textit{extra RRL}. These stars ($96$ RRL in $13$ galaxies) were included in our analysis of the spatial distribution of RRL in UFDs discussed in the next section.

\subsection{Spatial Distribution of the RRL} \label{subsec:spatial_dist}

Of the examples shown in Fig. \ref{fig:SpatialProjection}, we note that all of the four galaxies have RRL located at distances greater than $4\, R_h$. In this section we analize how frequent is to have such extended populations in UFDs. We arbitrarily adopted $4\, R_h$ as the distance to consider if a UFD has an extended population or not. Our motivation to choose this specific value is driven by the fact that almost all of the light of the galaxies is expected to be contained inside a $4\, R_h$ radius, so the stellar material found outside of $4\, R_h$ can be considered as stellar halo, or stellar debris. For comparison, \cite{Waller2023} analizes the outskirts of $3$ UFDs and they choose distances larger than $\sim 2\, R_h$ from the center of their host galaxies as evidence of extended populations. Here we are more conservative and push faraway the distance to be considered external population.

In order to understand how common it is to find RRL stars at large angular separations from the host galaxy, we calculated their elliptical separations, defined as half the geometrical constant of the ellipse defining the morphology of the UFD at the location of each individual RRL (Eq. \ref{eq:ellsep}): 

%\begin{equation}
\begin{multline} 
\label{eq:ellsep}
    S_{\rm ell} = \{ [ x \cdot \cos(90\degr - {\rm PA}) + y\cdot \sin(90\degr - {\rm PA}) ]^2 + \\
          [-x\cdot \sin(90\degr - {\rm PA}) + y\cdot \cos(90\degr - {\rm PA}) / (1-\epsilon) ]^2 \}^{\frac{1}{2}}
\end{multline}
%\end{equation}

\noindent
where (x,y) are the planar coordinates of the RRL, centered in their respective UFDs, and $\epsilon$ and PA are the galaxy's ellipticity and position angle respectively. The latter values were taken from Table \ref{table:dataUFDs}. Each elliptical separation was normalized by the $R_h$ of each galaxy. These normalized values for each RRL are listed in Table \ref{table:RRLdata} (column $15$). 

\begin{figure*}
    \begin{centering}
    \includegraphics[width=1\columnwidth]{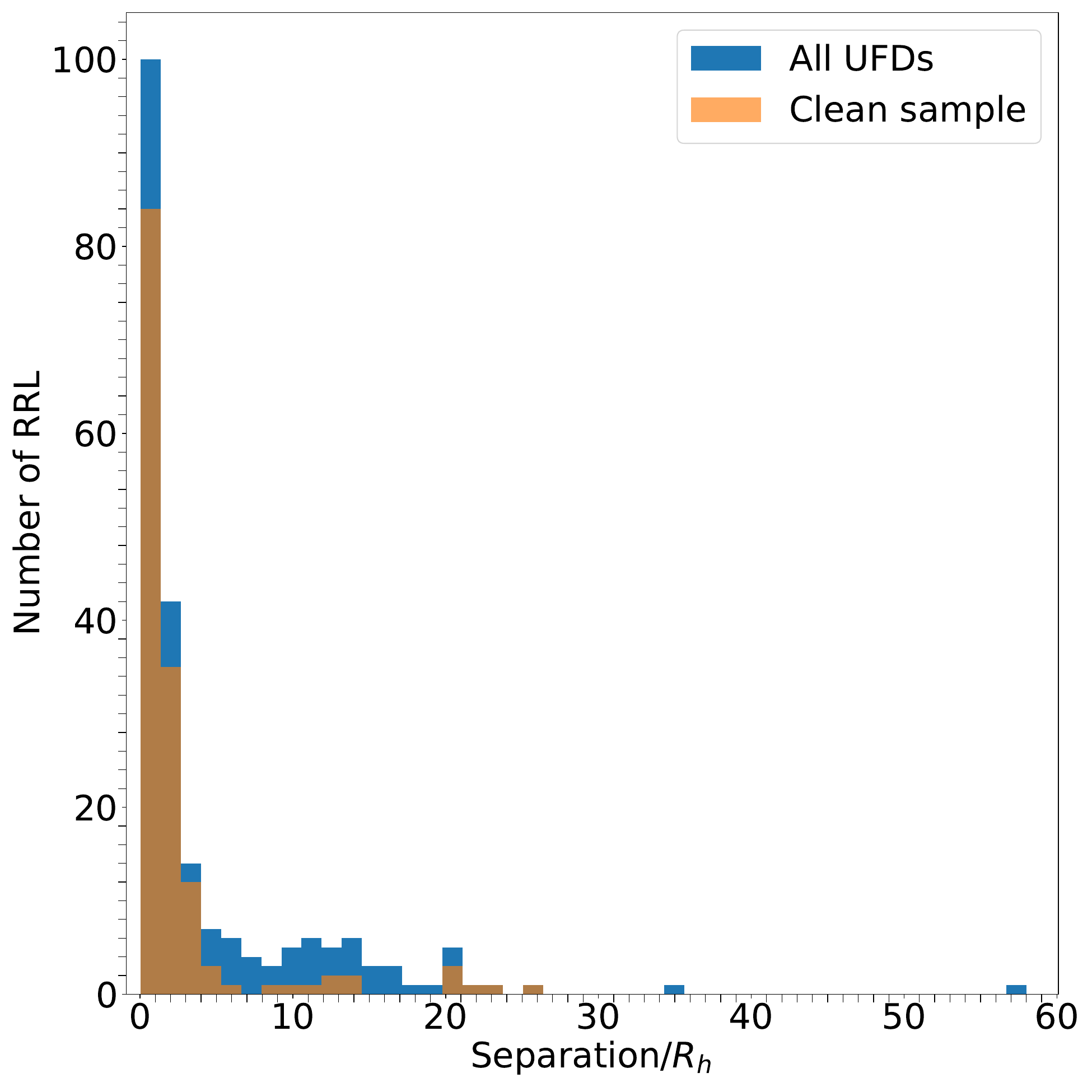}
    \includegraphics[width=1\columnwidth]{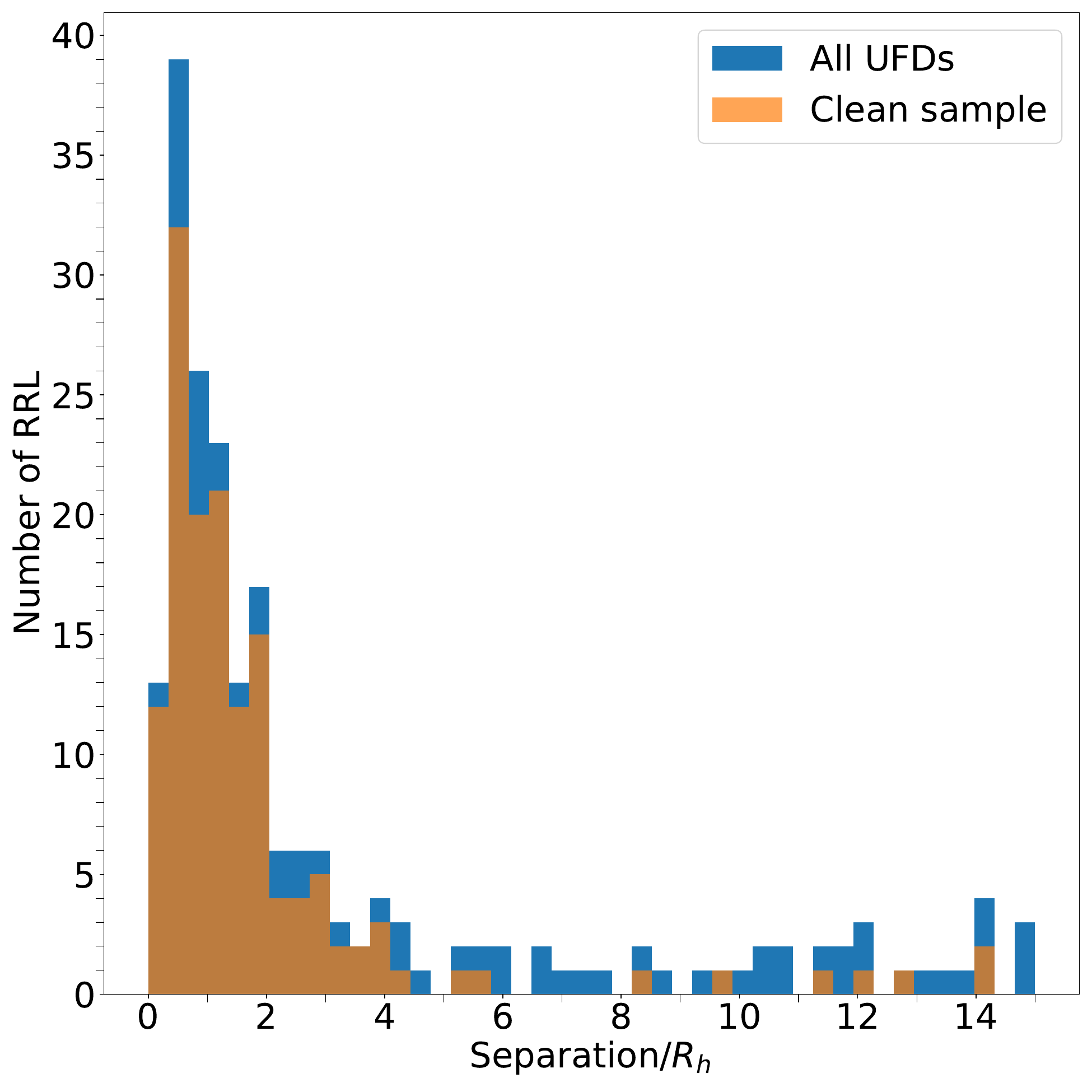}
    \caption{\textit{Left panel}: distribution of the elliptical separation of RRL from their respective host galaxies, using their $R_h$ as a normalization. In blue we show the distribution of the normalized elliptical separation of all the RRL (those found in this work plus those added as \textit{extra RRL} as mentioned in Sec. \ref{subsec:UFDnoRRL}), and in orange, we highlight the distribution of the RRL that belong to our clean sample of UFDs plus the extra galaxies mentioned in Sec. \ref{subsec:UFDnoRRL}. \label{fig:SepHist}. \textit{Right panel}: zoom of the left panel in the range $S_{ell}<15\arcmin$.}
    \end{centering}
\end{figure*}

In the left panel of Fig. \ref{fig:SepHist} we show the distribution of the normalized elliptical separations for all RRL associated with a UFD, including the ones in the \textit{extra RRL} list described in the previous section, leaving us with a total of $216$ RRL. The histogram shows that RRL have elliptical separations from the center of their host galaxies ranging from $0$ to almost $60\, R_h$. Although our angular distance main criteria (Section \ref{subsec:angsep_select}) was to select stars only within a circle of $15\; R_h$, for very small galaxies we allowed a search area of $1\degr$, which for tiny galaxies such as DELVE 2 means several tens of $R_h$, and that is the reason for finding some very distant RRL stars in our sample. As we discussed in Sec. \ref{subsec:contamination}, Bootes I, Bootes III, Tucana IV, DELVE 2, and Grus II  are special cases because there are concerns that contamination may be affecting the sample of RRL we identified as members.  We excluded the RRL in those galaxies to build a ``clean" sample. In Grus II, we actually know which stars are the contaminants (from the Chenab/Orphan stream in that case) since the magnitude distribution is bimodal; thus, we excluded only the 4 brighter RRL originally associated with this galaxy. In consequence, our clean sample is made of $142$ RRL. A histogram of the distribution of the normalized $S_{ell}$ of the clean sample is also shown in Fig. \ref{fig:SepHist}. It is clear that although many of the RRL far away from the center of their UFDs likely come from the galaxies affected by contamination, we still find several distant RRL in the clean sample of UFDs.

A zoom of the same histogram, including only  $S_{ell}<15 R_h$, is shown in the right panel of Fig. \ref{fig:SepHist}. The histograms show that the majority of the RRL are located in the central regions of their galaxies, but the number of RRL located in the outer regions is not negligible. Out of the total of $216$ RRL, there are $127$ RRL ($\sim 59\%$) inside $2\, R_h$ and $156$ RRL ($\sim 72\%$) inside $4\, R_h$. On the other hand, there are as many as $60$ RRL ($\sim 28\%$) that are located far away from their host galaxies ($S_{ell}\geq 4\, R_h$), $17$ of those belonging to galaxies of the clean sample ($8\%$ of the clean sample), i.e. galaxies where contamination by other stellar systems is not suspected. 
%$15$ of those in galaxies in the clean sample ($18\%$ of the clean sample), i.e. galaxies where contamination by other stellar systems is not suspected. 

\begin{figure}[!h]
    \begin{centering}
    \includegraphics[width=\columnwidth]{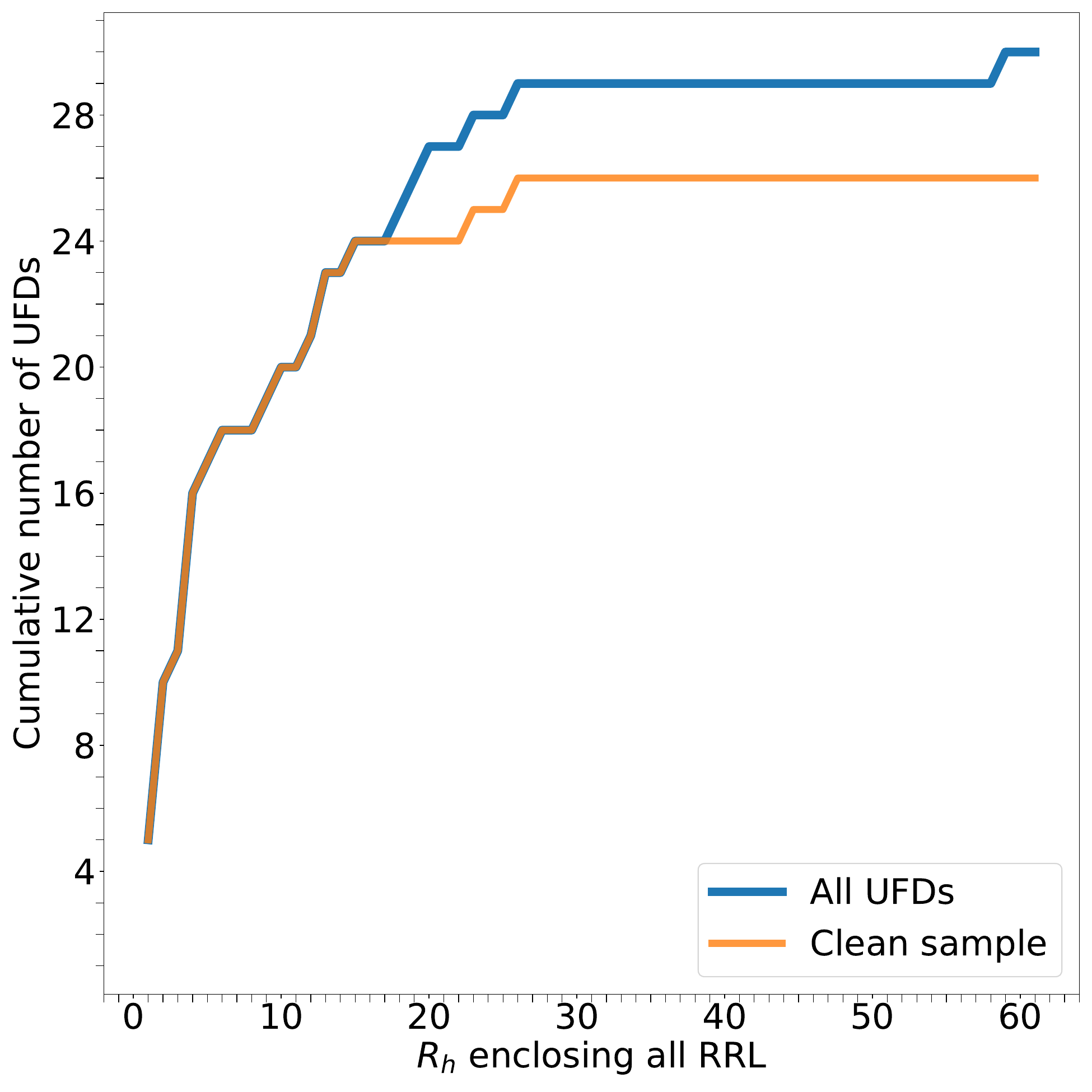}
    \caption{Cumulative number of galaxies that have all of their RRL enclosed within certain times their respective $R_h$.}
    \label{fig:GalHist}
    \end{centering}
\end{figure}

In Fig. \ref{fig:GalHist} we show the cumulative number of UFDs with all of their RRL enclosed within certain $S_{ell}$. We find that $16$ out of the $30$ UFDs with RRL ($\sim 53\%$) have all of their RRL within $4\, R_h$. On the other hand, there are $14$ UFDs ($\sim 47\%$) that have at least one RRL located at a distance $\geq 4\, R_h$, and $10$ UFDs ($\sim 33 \%$) that have at least one located at a distance farther than $10\, R_h$. If we use only the clean sample, then there are 10 UFDs ($\sim 33\%$) that have at least one RRL located at a distance $\geq 4\, R_h$, and $6$ UFDs ($\sim 20 \%$) that have at least one RRL located at a distance farther than $10\, R_h$. The galaxies in the clean sample with RRL at $S_{ell}\geq 4\, R_h$ are Tucana II, Tucana III, Grus II, Sagittarius II, Pegasus IV, Hercules, Ursa Major I, Centaurus I, Leo IV, and Eridanus II. We notice also that Pegasus III, Ursa Major II, Coma Berenices and Hydrus I have at least one RRL outside $S_{ell}>3.5\, R_h$. These numbers suggest that the existence of extended stellar populations is not uncommon in UFDs.

\section{Discussion and Conclusions} \label{sec:conclusions} 

We analyzed a sample of $45$ ultra-faint dwarf galaxies of the Local Group (see Table \ref{table:dataUFDs}). We used data from the Gaia DR3, DES, ZTF and PS1 surveys to search for RRL in these galaxies. Our main goal was to characterize the spatial distribution of RRL within each galaxy and to study how frequent it is to have extended stellar populations in UFDs. We found a total amount of $120$ RRL belonging to $21$ different UFDs (Table \ref{table:RRLdata}). 

Some of these stars are reported for the first time. For Hydrus I we report one new RRL, bringing the total of known RRL to $5$ after adding those previously reported by \cite{koposov18} and \cite{vivas20}. For Ursa Major II, besides the $4$ previously known RRL \citep{dallora12, vivas20}, we report $2$ new RRL. For Grus II we report one new RRL that adds to the one previously reported by \cite{martinez19}. For Tucana II there were $3$ RRL previously known \citep{vivas20} and we report one new RRL. For Ursa Major I, in addition to the $7$ RRL reported in \cite{garofalo13}, we found one new RRL that is located outside the field of view studied by those authors. For Eridanus II, besides the $67$ RRL previously found by \cite{MartinezVazquez2021a}, we report one new RRL with a location that is outside the region studied in that work. In regard to Bootes I and Bootes III, we report various new RRL in both galaxies but we recognize that these UFDs (in particular Bootes I) are affected by contamination from the Sgr stream. More studies involving the stars' metallicities and radial velocities are needed to determine the association of these RRL with either the galaxies or the stream. In the case of Segue I, the RRL found in this work is the same one reported by \cite{simon11}, but here we confirm for the first time its period using PS1 data. On the other hand, in this work we withdraw the previous claim that Eridanus III and Reticulum III had one RRL each \citep{vivas20}. In both cases, neither one of the RRL was found in Gaia DR3's RRL catalog \citep{Clementini2022}, suspecting a misclassification in the previous DR2.

\begin{figure}[!h]
\plotone{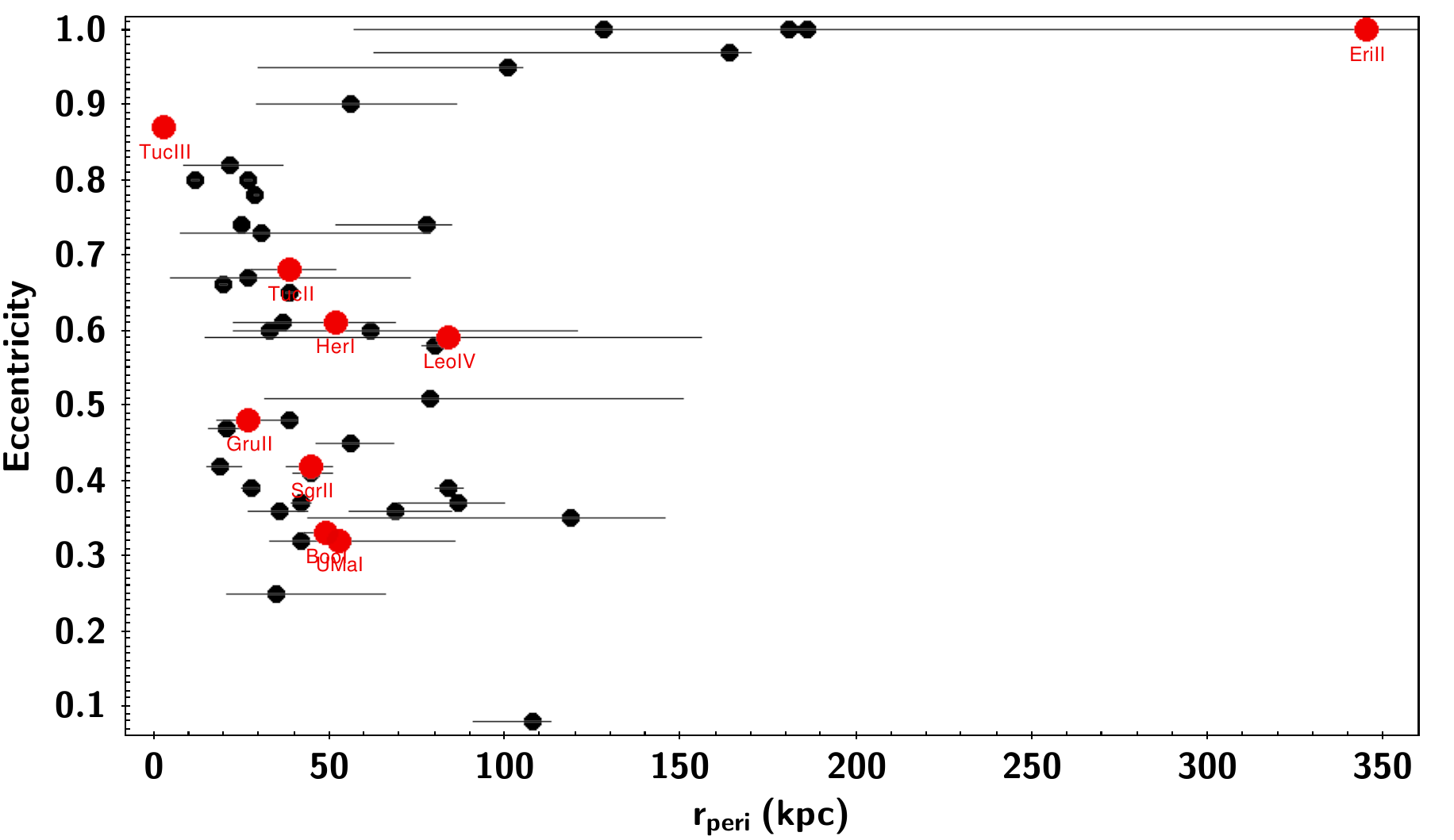}
\caption{Eccentricity and pericenter distance of the orbits of UFD galaxies calculated by \citet{Li2021} with Gaia EDR3 proper motions and assuming a Milky Way potential that follows an Einasto profile with high mass (their model PE$_{\rm HM}$). Red symbols indicate UFD galaxies with extended populations as identified in this work. Neither Pegasus IV nor Centaurus I are included in this plot since they were not available at the time of \cite{Li2021} orbit calculations.}
\label{fig:peri}
\end{figure}

Using Eq. \ref{eq:ellsep} we computed the elliptical separation between these selected RRL and their respective host galaxies, and normalized them by their respective $R_h$. We found that $14$ UFDs have RRL located at distances from their center that are $\geq 4\, R_h$, and we are certain that $10$ of them do not have contamination problems. Therefore, \textit{we conclude that UFDs can have extended populations}. These $14$ UFDs with RRL in their outskirts are located at heliocentric distances ranging from $\sim 23$ kpc to $370$ kpc and have absolute magnitudes ranging from $-1.3$ to $-7.1$, so there does not seem to be a preference neither in these galaxies' distance distribution from the Milky Way nor in their absolute magnitude values. In Fig.~\ref{fig:peri} we show the distribution of the pericenter distance, $r_{\rm peri}$ and eccentricity of the orbits of the Milky Way UFD satellites as calculated by \citet{Li2021}. The orbits of UFDs with extended populations (red symbols) do not have particularly close approaches to the Milky Way, except for Tucana III, nor they have particularly eccentric orbits. Galaxies with small pericenter passages are strong candidates to suffer tidal disruptions that can be responsible for the presence of extended populations. However, there is a wide range of pericenter distances in the UFDs that we flagged as having extended populations, suggesting that other mechanisms besides interaction with the Milky Way may play a role in the formation of extended stellar populations around UFDs.

It is important to stress that the number of $10$ UFDs with external populations is a lower limit. Not only we have at least $2$ more UFDs (Bootes I and Bootes III) with candidates in the outskirts that need more follow-up in order to discard contamination by the Sgr stream, but also some of the most distant UFD galaxies (the ones outside Gaia's faint limit) may not have wide coverage for variable stars. Four additional UFDs would have extended populations if our arbitrary definition is lowered to $S_{ell}\geq 3.5 R_h$. It is also important to highlight that RRL are scarce in UFDs, with many of them hosting just a few (or none) of those variables \citep{martinez19,vivas20}. Thus, there may be cases in which the number of RRL in a UFD is just not enough to trace the population.

Recently, \cite{Waller2023} studied Ursa Major I, Bootes I and Coma Berenices  with the aim to identify potential members in the outskirts of these three UFDs using Gaia EDR3 data and the Gemini GRACES spectrograph. Comparing their work with our results for these same UFDs of our sample, we find them to be in good agreement. The authors report one star at $3.7\, R_h$ in the case of Ursa Major I, one star at $4\, R_h$ in Bootes I and two stars at $2.5 \, R_h$ for Coma Berenices, concluding that these galaxies have stars located in their outskirts. In this work we also identify stars in the outer regions of Ursa Major I and Bootes I. In the case of Ursa Major I, we found $1$ RRL in its outskirts according to our criterion (at $8.40\, R_h$), but we do also find one RRL at a distance similar to the one reported for the star studied in \cite{Waller2023}, located at $3.01\, R_h$. Regarding Bootes I, we found $12$ RRL that are possibly in its outer region (raging between at $4.0 - 17.8\, R_h$), but we bear in mind that this galaxy's RRL sample is probably contaminated by some stars that belong to the Sgr stream. \cite{Jensen2023} also identify $34$ stars that could be part of the outskirts of Bootes I using their own numerical algorithm (one of them being located at $9\, R_h$), but mention that a spectroscopic analysis would be useful to better understand this result. As for Coma Berenices, although we do not find RRL located at distances greater than $4\, R_h$, we found one RRL at $3.98\, R_h$ which is even farther away than those identified by \cite{Waller2023}. We also found agreement with previous studies of Tucana II in which stars up to $9\, R_h$ were identified in the external parts of the galaxy by \citet{Chiti2021}. Here we also found 2 distant RRL in Tucana II, at 5 and 12 $R_h$. A more detailed analysis of the RRL found in the outskirts of the UFDs could be made by studying their abundances, and radial velocities \citep[e.g.][]{Chiti2021}. These additional data would further corroborate their association with the UFDs and could help understand if they were formed in their central regions and then stripped away into farther distances or if an outside-in star formation scenario explains their current locations. 

These extended stellar populations can also be found in dSphs, such as Fornax \citep{Stringer2021, Qi2022}, Carina \citep{vivas13, Qi2022} and Ursa Minor \citep{Sestito2023a, Jensen2023}, so it is not a an exclusive feature of the UFDs.

Numerical simulations also provide information on the extended population of dwarf and UFD galaxies. The resolution of cosmological hydrodynamical simulations has been improving in the past years and this leads galaxy models towards being more fiducial at low-mass ranges. This type of analysis will allow us to study the evolution of the modelled galaxies by tracing them through different redshifts, which could help enlighten the mechanism responsible for driving the stars of these UFDs to such large distances from the center of their hosts.

This era of big surveys has facilitated the search for wider regions in the sky and has allowed us to extend to larger distances. The upcoming large scale surveys, such as the Vera Rubin Observatory and the Nancy Grace Roman Space Telescope, will reach farther regions when observing dwarf galaxies in the Local Group and will be able to get better and more accurate information regarding their true extent.

\begin{acknowledgments}

We thank the anonymous referee for useful comments and suggestions to this work. EAT acknowledges financial support from ANID ``Beca de Doctorado Nacional" 21220806. C.E.M.-V. is supported by the international Gemini Observatory, a program of NSF's NOIRLab, which is managed by the Association of Universities for Research in Astronomy (AURA) under a cooperative agreement with the National Science Foundation, on behalf of the Gemini partnership of Argentina, Brazil, Canada, Chile, the Republic of Korea, and the United States of America. This work has made use of data from the European Space Agency (ESA) mission {\it Gaia} (\url{https://www.cosmos.esa.int/gaia}), processed by the {\it Gaia} Data Processing and Analysis Consortium (DPAC, \url{https://www.cosmos.esa.int/web/gaia/dpac/consortium}). Funding for the DPAC has been provided by national institutions, in particular the institutions participating in the {\it Gaia} Multilateral Agreement.

\end{acknowledgments}

\vspace{5mm}
\facilities{Gaia, PS1, Blanco, PO:1.2m}

%\bibliography{UFD}{}
%\bibliographystyle{aasjournal}

\end{document}